%% file: main.tex
\documentclass[12pt]{article}
\usepackage[]{times} 
\usepackage{CJKutf8}

\usepackage[numbers]{natbib}
\makeatletter
\def\@mb@citenamelist{cite,citep,citet,citealp,citealt,citeplatex,citetalias}
\makeatother

\usepackage{multibib}  
\newcites{latex}{Appendix References}
\usepackage{geometry}
\usepackage{afterpage}
\usepackage{graphicx}
\graphicspath{ {./images/} }
\usepackage{float}
\usepackage{amsmath}
\usepackage{enumitem}
\usepackage{caption}
\usepackage{subcaption}
\usepackage{url} 
\usepackage{amsthm,letltxmacro}
\usepackage{textcomp}
\usepackage{gensymb} 

\usepackage{cleveref} 
\crefname{theor}{hypothesis}{hypotheses}

\usepackage{changepage}
\LetLtxMacro\oldproof\proof
\let\endoldproof\endproof
\renewenvironment{proof}[1][\proofname]
  {\begin{adjustwidth}{2em}{2em}
   \oldproof[#1]}
  {\endoldproof
   \end{adjustwidth}}

\newtheorem{definition}{Definition} 
\newtheorem{theor}{Hypothesis}

\crefname{quest}{question}{questions}
\newtheorem{quest}{Research Question}
\usepackage{amssymb}
\usepackage{pifont}
\usepackage{xparse}
\usepackage{adjustbox}
\usepackage{longtable}
\usepackage{multirow}
\usepackage{booktabs} 
\usepackage[flushleft]{threeparttable} 
\usepackage{threeparttablex}
\usepackage{array}
\usepackage{rotating}

\usepackage[bottom]{footmisc}

\newcolumntype{P}[2]{
  >{\begin{turn}{#1}\begin{minipage}{#2}\small\raggedright\hspace{0pt}}l
  <{\end{minipage}\end{turn}}
}
\usepackage{multicol}
\usepackage{xcolor,colortbl}

\definecolor{Gray}{gray}{0.85}
\definecolor{LightGray}{gray}{0.98}

\newcolumntype{a}{>{\columncolor{Gray}}c}
\newcolumntype{d}{>{\columncolor{LightGray}}c}

\newcommand\blfootnote[1]{
  \begingroup
  \renewcommand\thefootnote{}\footnote{#1}
  \addtocounter{footnote}{-1}
  \endgroup
}

\title{Experience of the COVID-19 pandemic in Wuhan leads to a lasting increase in social distancing}

\author{Darija Barak$^{\dagger,a}$, Edoardo Gallo$^{\dagger,a,*}$, Ke Rong$^{b,*}$, Ke Tang$^{b,*}$, Wei Du$^{c}$}

\begin{document}
\maketitle

\begin{abstract}
    On 11th Jan 2020, the first COVID-19 related death was confirmed in Wuhan, Hubei. The Chinese government responded to the outbreak with a lockdown that impacted most residents of Hubei province and lasted for almost three months. At the time, the lockdown was the strictest both within China and worldwide. Using an interactive web-based experiment conducted half a year after the lockdown with participants from 11 Chinese provinces, we investigate the behavioral effects of this `shock' event experienced by the population of Hubei. We find that both one's place of residence and the strictness of lockdown measures in their province are robust predictors of individual social distancing behavior.  Further, we observe that informational messages are effective at increasing compliance with social distancing throughout China, whereas fines for noncompliance work better within Hubei province relative to the rest of the country. We also report that residents of Hubei increase their propensity to social distance when exposed to social environments characterized by the presence of a superspreader, while the effect is not present outside of the province. Our results appear to be specific to the context of COVID-19, and are not explained by general differences in risk attitudes and social preferences.
\end{abstract}



\blfootnote{We are grateful to Alastair Langtry and Ilia Shumailov for helpful discussions. We acknowledge support from the National Natural Science Foundation of China (grant no.71872098 for KR and grant no. 72192802 for KT), the Beijing Social Science Fund (KR grant no.21DTR051), the Guoqiang Institute of Tsinghua University, Cambridge Humanities Research Grant Scheme (EG), Economic and Social Research Council and Trinity College Cambridge (DB). This work is the sole responsibility of the authors, and does not necessarily represent the official views of any of the funding agencies.}
\blfootnote{$a$ University of Cambridge, Faculty of Economics, Sidgwick Avenue, Cambridge, UK.}
\blfootnote{$b$ Institute of Economics,  School of Social Sciences, Tsinghua University,  Haidian District, Beijing, 100084, China}
\blfootnote{$c$ School of Economics, Anhui University of Finance and Economics, Bengbu, Anhui, 233030, China}
\blfootnote{\noindent 
$\dagger$ DB and EG contributed equally to this work.}
\blfootnote{\noindent 
$*$ To whom correspondence should be addressed: EG (edo@econ.cam.ac.uk), KR (r@mail.tsinghua.edu.cn), KT (ketang@tsinghua.edu.cn).}

\clearpage
\input{sections/main_text.tex}

\clearpage
\bibliographystyle{ieeetr}
\bibliography{main_bib.bib}

\renewcommand{\thefigure}{A\arabic{figure}}
\renewcommand{\thetable}{A\arabic{table}}
\setcounter{figure}{0}

\clearpage
\appendix

\section{Supplementary Information}
This Appendix contains supplementary information. Section \ref{sec:model} specifies our theoretical framework, calibration for the experiment, and the set of hypotheses we investigate. Section \ref{sec:methods_collection} describes our methodology of data collection, including recruitment and the actual experiment. Section \ref{sec:dataset} presents our dataset, and Section \ref{sec:methods_analysis} explains our data analysis methodology. Section \ref{sec:results} presents our key results, while Section \ref{sec:robustness} summarizes our key robustness checks.

\subsection{Theory and hypotheses}
\label{sec:model}
\input{sections/theory.tex}

\subsection{Methodology: data collection} \label{sec:methods_collection}
\input{sections/methods_collection.tex}

\subsection{Dataset} \label{sec:dataset}
\input{sections/dataset.tex}

\subsection{Methodology: data analysis} \label{sec:methods_analysis}
\input{sections/methods_analysis.tex}

\subsection{Results} \label{sec:results}
\input{sections/results.tex}

\subsection{Robustness checks} \label{sec:robustness}
\input{sections/further_analysis.tex}

\clearpage
\bibliographystylelatex{ieeetr}  
\bibliographylatex{appendix_bib.bib} 

\clearpage
\section{Experimental Instructions} \label{sec:instructions}
\setlength{\parskip}{6pt}
\setlength\parindent{0pt}
\renewcommand{\thefigure}{B\arabic{figure}}
\renewcommand{\thetable}{B\arabic{table}}
\setcounter{figure}{0}
\input{sections/instructions.tex}

\end{document}

%% file: sections/main_text.tex
\section{Introduction}

The COVID-19 pandemic has brought the most significant and devastating global disruption since World War II with an estimated 5.5 million deaths worldwide \cite{W_2021, WB_2020, 10.1093/ije/dyab207}. Most countries implemented drastic lockdown policies to minimize infection levels, prevent healthcare systems from being overwhelmed, and reduce the number of deaths \cite{H_2021, VKB_2020, Braunereabd9338, A_2021}. The first COVID-19 related lockdown started on 23rd Jan 2020 in Wuhan (Hubei) and for the subsequent 3 months the measures taken in Hubei were the strictest both within China and worldwide. In particular, according to the Oxford COVID-19 Government Response Tracker (OxCGRT), the average government response index value for Hubei in this period was 75.8 (with 0 being no measures and 100 being the maximum) \cite{H_2021}. Meanwhile, China as a whole scored 58.5, and the closest scoring countries -- Italy and Mongolia -- had an average index of 53.0 and 49.9 respectively. It is estimated that the measures implemented by the Chinese government have potentially prevented 100,000s of COVID-19 infections \cite{HT_2020}, and possibly contributed significantly to public health in China overall \cite{QZZ_2021}.

The policies aimed at containing the spread of the pandemic have had a profound impact. Research into the impact of lockdowns and other Non-Pharmaceutical Interventions (NPIs) has documented a deterioration in physical and mental health in China \cite{wang2020immediate,wang2020longitudinal}, as well as other countries \cite{le2020anxiety, tran2020impact, wang2020association}. Recent evidence shows there are mental health and burnout effects associated with the Zero-COVID policies \cite{lau2022covid}. However, to date, little attention has been paid to the effects of the pandemic on human behavior. Our paper addresses this gap.

Previous research suggests that shock events and/or drastic institutional interventions can have a long-lasting impact on behavior \cite{AJR_2005, PLR_1993}. For example, colonial conscription rules in 16th century Bolivia and Peru led to differences in household consumption that survive to this day \cite{D_2010}. More recently, the 2004 tsunami in Thailand led to significant increases in prosocial behavior, risk aversion, and impatience in rural areas \cite{CJK_2017}. The sudden and drastic nature of the COVID-19 outbreak and associated lockdown policies in Hubei compared to the rest of China means they may have persistent effects on the behavior of Hubei’s residents in the short and medium term, especially when it comes to reacting to post-lockdown policies to contain the pandemic. The sudden and drastic nature of the COVID-19 outbreak and associated lockdown policies in Hubei compared to the rest of China means they may have persistent effects on the behavior of Hubei’s residents in the short and medium term, especially when it comes to reacting to post-lockdown policies to contain the pandemic.

Using an interactive web-based experiment conducted half a year after the end of the lockdown, we show that Hubei residents behave differently compared to inhabitants of other provinces in China in terms of social distancing, receptiveness to COVID-19 policies, and when exposed to a superspreader environment. In particular, we estimate that every extra 1,000km between Wuhan and one’s place of residence contributes to a 7\% decrease in social distancing. Using OxCGRT, we show that an increase in the harshness of lockdown measures is associated with an increase in distancing. 

The differences in social distancing behavior between residents within and outside Hubei may translate into differences in social distancing policies effectiveness. We examine the effect of soft and hard policy interventions to promote social distancing. The hard policy intervention -- a fine -- increases social distancing only in Hubei, while the soft intervention -- an informational message or “nudge” -- increases social distancing both within and outside Hubei. Finally, Hubei-based participants practice more social distancing in a social environment with a superspreader. Using data from incentivized preference elicitation tasks, we find that the observed differences in behavior between Hubei residents and those from the rest of China are not explainable by general differences in preferences.

\section{Experimental design}

\begin{figure}[h]
    \centering
    \includegraphics[width=0.6\linewidth]{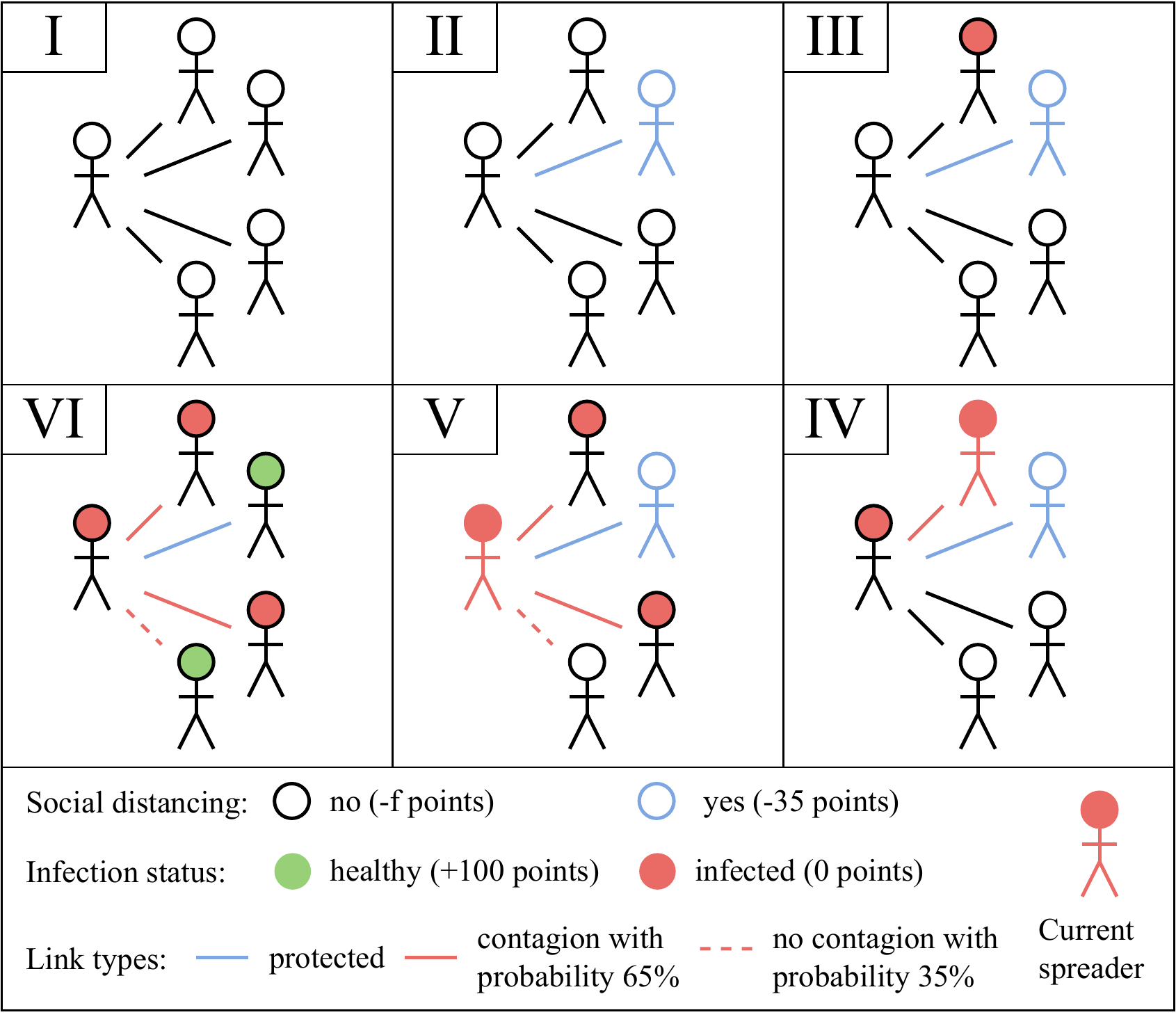}
    \caption{Flow of a typical round of baseline and intervention. In the experiment, we use the following parameterization: $f = 0$ points in baseline and in nudge intervention, and $15$ points in fine intervention. Final payoffs for the round are a combination of individual social distancing choice and infection status. For example, a participant who practices social distancing and is healthy, receives $(-35+100) = 65$ points. In the figure, the chosen social environment is the superspreader. In the experiment, half of the treatments had superspreader environment while the other half had a homogeneous environment.}
    \label{fig:0}
\end{figure}

\noindent \textbf{The game.} Figure \ref{fig:0} presents the flow of a typical round of the experiment. Participants are randomly assigned to groups of five that stay the same throughout their involvement in the study. Within the group, in each round they are randomly assigned to five positions within the social structure -- nodes on a network -- as shown in Panel I). In every round each participant has to make a binary decision of whether or not to practice social distancing. Each participant has to privately decide whether to practice social distancing at a cost of 35 points. In the example in Panel II, the participant color-coded in blue is the only one who chose to practice social distancing. Once decisions are made, the computer picks one subject to be potentially infected by COVID-19 uniformly at random. If this patient zero subject practices distancing, she becomes infected with probability 50\%. If patient zero does not practice distancing, infection happens for sure. COVID-19 then spreads from infected to healthy participants who do not practice social distancing with a commonly known probability of 65\%. Note that those who practice social distancing cannot (a) infect others or (b) become infected through this contagious process. 

An example of such contagious process is in panels III-V of Figure \ref{fig:0}. Panel III shows patient zero color coded in red. Given that patient zero chose not to practice distancing, there is a $65\%$ chance that the participant in the superspreader position, who does not practice distancing either, gets infected. Panel IV shows the case when the participant in the superspreader position gets infected, and can therefore spread COVID-19 to all other participants. Finally, Panel V shows the instance when infection occurs for one out of the two remaining participants who do not practice distancing and are connected to the superspreader. Panel VI shows the final outcome of the spreading process with three participants infected and two remaining healthy. 

At the end of the round, healthy participants receive 100 points while those infected get 0 points, minus costs of social distancing if applicable. For example in panel VI of Figure \ref{fig:0}, three participants receive a payoff of 0, one gets 65 points, and another one -- 100 points. Throughout both instructions and the experiment, participants are primed to think about COVID-19. For full details on the instructions and the experimental interface, consult the Supplementary Information (SI).\\

\noindent  \textbf{Treatments.} Participants play 20 rounds of the above social distancing game, which constitute the baseline part of the experiment. After these 20 rounds, they are treated with one of the policy interventions. The soft policy intervention is an informational message or \emph{nudge} -- participants must watch a 3-minute video which explains how failure to practice distancing can harm others. The hard policy intervention is the introduction of a fine of 15 points for everyone who does not practice social distancing in a round of the game. Participants play another 20 rounds of the social distancing game under either the nudge or fine policy intervention. Note that the payoff structure remains unchanged in the nudge treatments (Figure \ref{fig:0} with $f = 0$ points), while in the fine treatments subjects receive the fine in every round where they do not practice distancing irrespective of their health status (Figure \ref{fig:0} with $f = 15$ points).

A second treatment dimension is the social environment. Participants are randomly assigned to either a \emph{homogeneous} or a \emph{superspreader} environment (as in Figure \ref{fig:0}), which stays the same throughout the 40 rounds of the experiment. In the homogeneous case, everyone is connected to everyone else in the group so an infected participant can spread COVID-19 to any other healthy participant that does not practice distancing. In the superspreader case, one participant is connected to all the others, and there are no other connections in the group. This means that any spread of infection beyond patient zero must involve the central participant either as the spreader or the recipient. A defining feature of COVID-19 is the crucial role of superspreaders in the diffusion of the disease \cite{AWWLTCLC_2020, LGTBNL_2020}. For respiratory syndromes, an important determinant of being a superspreader is biological \cite{lloyd2005superspreading, edwards2021exhaled}, something that is typically unknown to the individual and outside the scope of this study. Another determinant is, however, the centrality of the individual in terms of the structure of social interactions -- this is typically common knowledge and varies widely across individuals in most environments \cite{B_1999}. This treatment dimension, therefore, allows an investigation of how a social environment with a superspreader affects the propensity to social distance.

Using a full-factorial $2\times2$ design, we therefore obtain four treatments. As standard in the experimental literature, subjects were randomly assigned to treatments, so the effect of our treatment variables is causal. We collect at least 10 groups of 5 subjects for each of those treatments. Additionally, to investigate the impact of Hubei residence on behavior, we ran these four treatments separately in Hubei and rest of China.  We summarize the details of our dataset next. 

\section{Dataset}

\begin{figure}[t]
    \centering
    \includegraphics[width=0.6\linewidth]{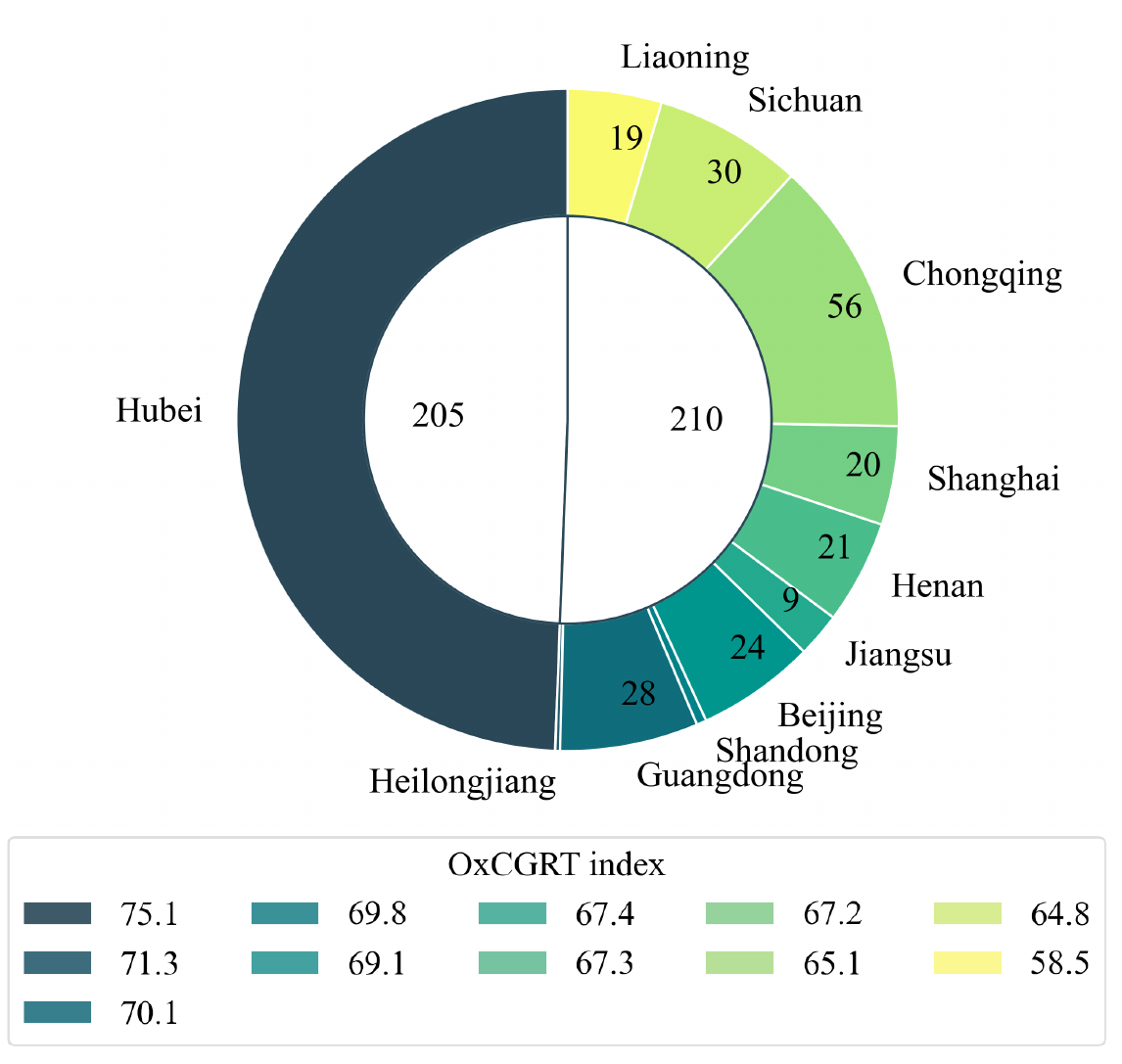}
    \caption{Distribution of subjects from different Chinese provinces. Color coding indicates the average values of the OxCGRT index for the provinces. }
    \label{fig:1}
\end{figure}

Using a local recruitment company, we sourced 415 participants from 11 Chinese provinces. Figure \ref{fig:1} displays the proportion of participants from each province. In the final sample, 205 subjects (41 groups) are from Hubei province and 210 (42 groups) from the rest of China. We verify the place of residence using (1) self-reported data from the recruitment survey, (2) data from the survey company, and (3) IPs of subjects collected when completing the experiment. Despite the possibility of selection bias, we obtain a diverse sample in terms of age and gender. In particular, 47.3\% of our sample is female and the mean age is 35 years (s.d. 10 years). Figure \ref{fig:1} also shows the average stringency of lockdown measures in the 11 provinces in our sample over the period of the Hubei lockdown as measured by the 0--100 scale of the OxCGTR index. Note that throughout the paper we focus on the difference between Hubei and other 10 provinces which were under more moderate lockdown measures. Further note that since we have between 1 and 56 subjects from each of the other 10 provinces, we cannot comment on the differences between these provinces.

\section{Results}

We analyze the data using a linear probability model, where the individual decision to practice social distancing is the dependent variable (binary), and the controls are (1) our treatments, (2) a set of demographic variables and variables for personal preferences, plus (3) a variable capturing the distance of one’s place of residence from Wuhan. Estimated coefficients from this model are in M1 Table \ref{tab:1}.\\

\input{tables/main_text_table}

\noindent \textbf{Hubei province versus the rest of China. }Our first finding is that the individual propensity to practice social distancing in our experiment is inversely related to distance from Wuhan to one’s place of residence. We estimate that every extra 1000km between Wuhan and one’s place of residence contributes a 7 percentage point decrease in the probability of social distancing. In practical terms, this suggests that the individual propensity to do social distancing of residents of Chongqing, which is approximately 723km away from Wuhan, to be 5 percentage points less than that of Wuhan residents. The effect is statistically significant in all our specifications and robust (M1, $p = 0.02$). Replacing the distance variable with a dummy equal to one for Hubei subjects (M2 Table \ref{tab:1}), we estimate that the probability of social distancing is 8.5 percentage points higher in Hubei province than outside of it (M2, $p = 0.006$).

We hypothesize that distance from Wuhan captures heterogeneity in the harshness of the lockdown policy experienced by people from different parts of China. While tight COVID-19 related restrictions were generally experienced throughout the world in early 2020, Wuhan was the first to go under total lockdown for a nearly 3-months period together with its 11 million residents \cite{NYT_2021a}. According to the OxCGRT, Hubei province has spent the whole 23rd Jan-2nd May 2020 period in a a very strict lockdown, whereas other provinces (with the exception of Heilongjiang) mostly experienced more moderate measures \cite{H_2021}. To test this hypothesis, we use data from the OxCGRT, which tracks harshness of government response to the COVID-19 pandemic globally. We focus on the 23rd Jan-2nd May 2020 period, and calculate the average of the overall government response index for each of the provinces in our sample. In this way we obtain a single index on a $[0,100]$ scale. The correlation between distance from Wuhan and this index for the 20 cities in our sample is -0.6683 (t-test, $p = 0.001$). Replacing distance of one's place of residence from Wuhan by the index, we estimate that a 1 point increase in the response index corresponds to a 0.75 percentage point increase in individual propensity to do social distancing in our experiment, which is significant in all specifications considered (M3 Table \ref{tab:1}, $p = 0.03$).

An important caveat is that the lockdown measures were not imposed randomly -- stricter measures were put in place in provinces with more severe COVID-19 outbreaks. The Hubei outbreak was by far the largest in China, with over 68,300 infections and about 4,500 COVID-19 related deaths confirmed at the time of writing \cite{ST_2021}. Guangdong -- the second province by the size of the outbreak -- recorded approximately 3,300 cases and 8 deaths. Indeed, the correlation between measures of harshness and number of confirmed cases for 11 provinces in our sample is 0.6835 (t-test, $p = 0.001$). An alternative interpretation of our results is, therefore, that the experience of the severity of the outbreak, rather than the lockdown measures, is the primary driver of the behavioral differential. While we are unable to differentiate between these alternative channels, the central message remains that what Hubei participants experienced caused a lasting change in their social distancing behavior.

The association between social distancing behavior and the severity of the COVID-19 pandemic experience raises two natural questions. The first one is whether there is a difference between participants from Hubei, who lived through a larger outbreak and harsher lockdown, and those from the rest of China in terms of their behavioral responses to our treatment variables -- fine/nudge policy interventions and homogeneous/superspreader social environments. The second one is whether the association is driven by specific demographic characteristics and/or risk/social preferences. In order to investigate this, we repeat our core analysis but interact every variable with a dummy equal to one if a subject is from the Hubei province and zero otherwise. This is equivalent to fitting the model separately on the two datasets. The results are reported below, and the full table with coefficients can be found in the SI.\\

\begin{figure}[t]
\centering
\hspace{-1.5cm}
\begin{minipage}{.47\linewidth}
\centering
\subfloat[]{\label{fig:2a}\includegraphics[scale=.55]{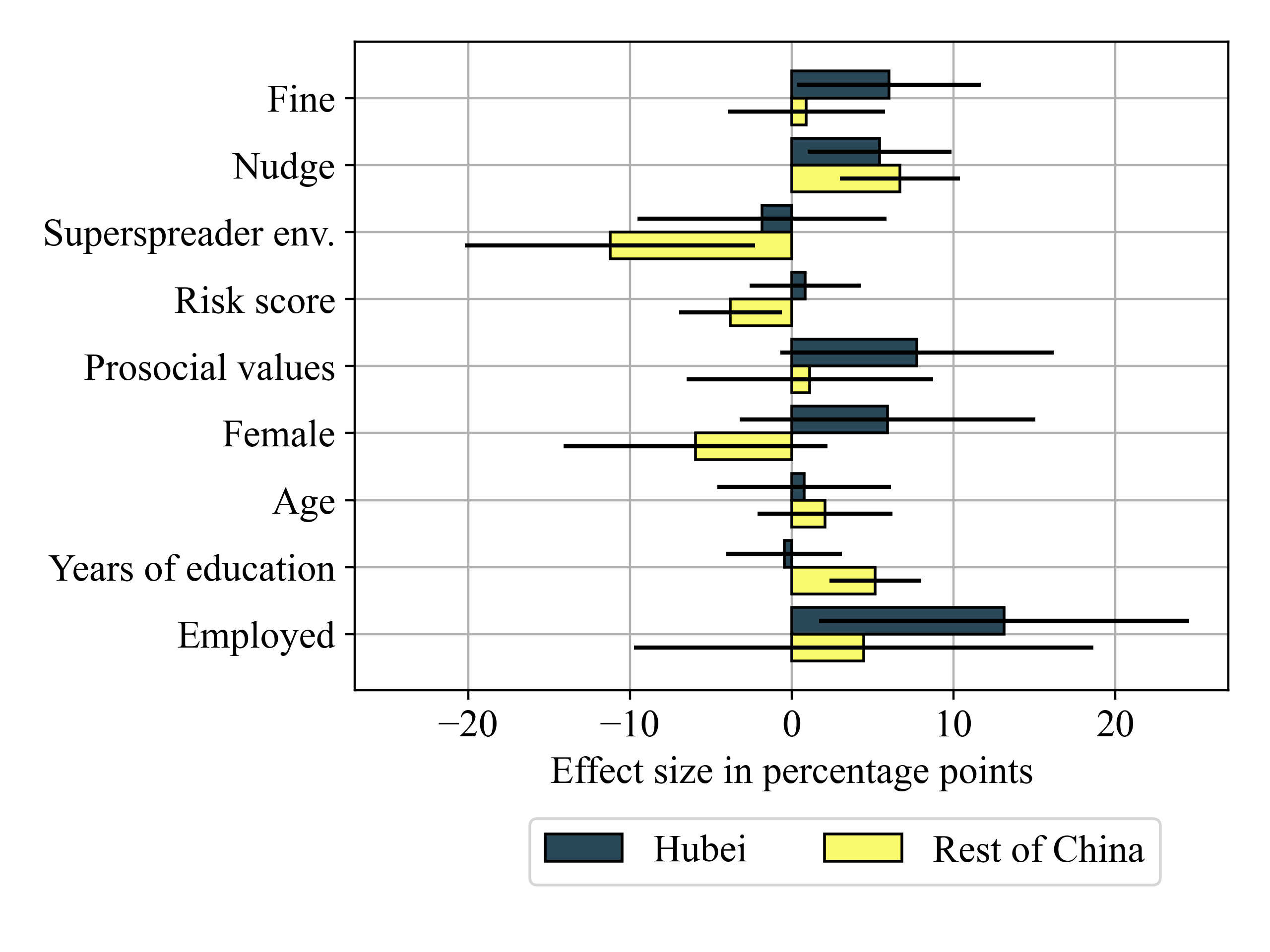}}
\end{minipage}
\hspace{0.5cm}
\begin{minipage}{.47\linewidth}
\centering
\subfloat[]{\label{fig:2b}\includegraphics[scale=.55]{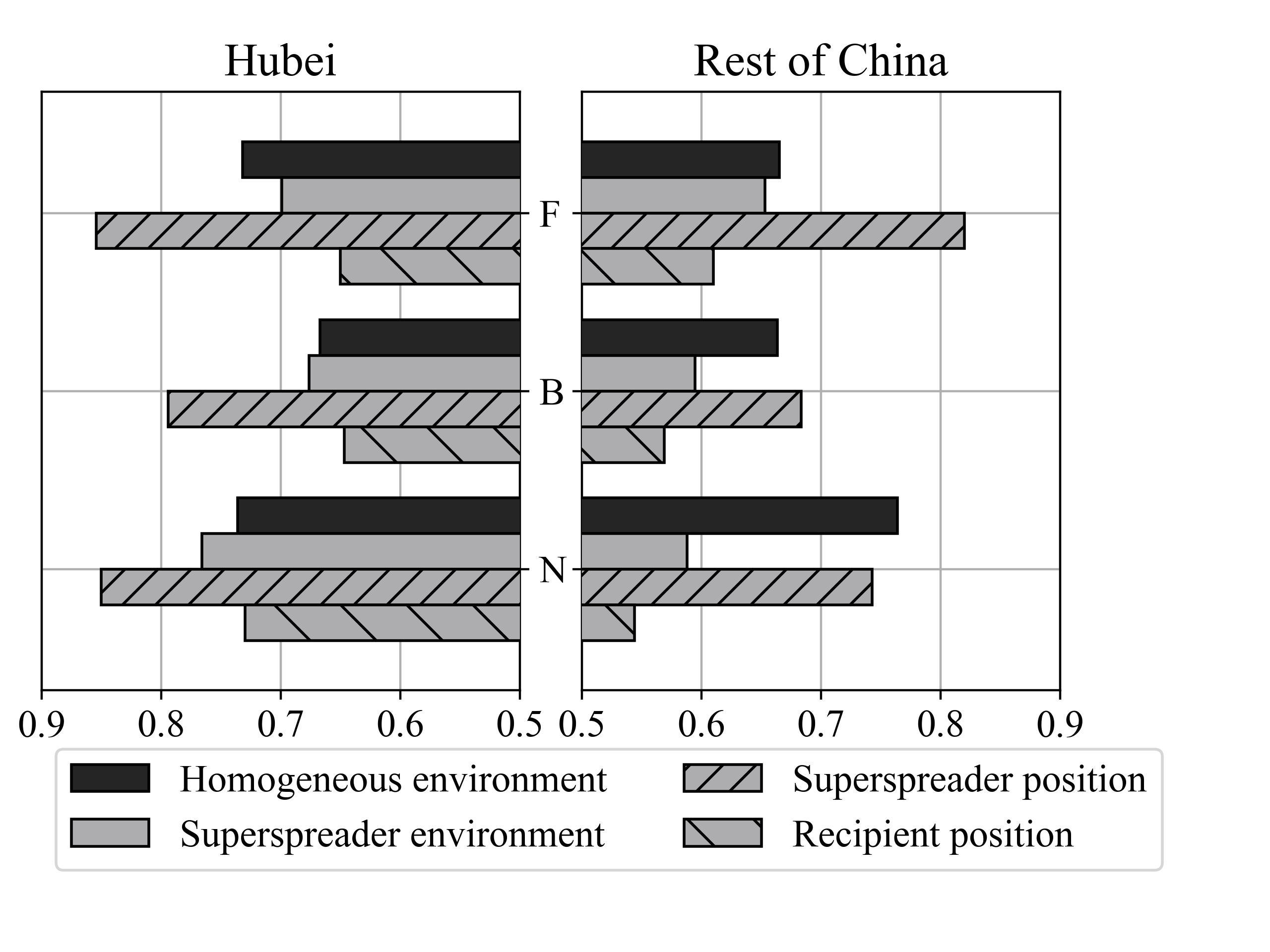}}
\end{minipage}
\caption{\textbf{(a)} Estimated effect size on the probability of social distancing for treatment variables and selected controls, split by Hubei and rest of China. Note that: (1) risk score estimates are reported for an extra 25 points, (2) Prosocial=1 for subjects with prosocial values as classified by the SVO scale, (3) Female=1 for female subjects, (4) age estimates are reported for an extra 10 years, (5) Employed=1 if subject is either employed or runs their own business. \textbf{(b)} Probability of doing social distancing separately for Hubei and rest of China in baseline part of the experiment (B), and under fine (F) and nudge (N) interventions split by social environment and positions in the superspreader environment.}
\label{fig:effect_size_v2}
\end{figure}

\noindent  \textbf{Response to intervention.} The hard policy intervention of introducing a fine increases the propensity to social distance in Hubei, but not in the rest of China. As shown in Figure \ref{fig:2a}, the fine leads to a significant 6.0 percentage points increase in individual propensity to social distance in Hubei (t-test, $p = 0.04$), while outside of Hubei the estimated effect is only 0.9 percentage point and not significant (t-test, $p = 0.7$). The difference between the two effects is not statistically significant (t-test, p = $0.2$).

In contrast, the soft policy intervention (the nudge) increases the propensity to social distance throughout China. As shown in Figure \ref{fig:2a}, the estimated size of the effect is 5.4 percentage points (t-test, $p = 0.02$) in Hubei province and 6.7 percentage points (t-test, $p = 0.0004$) in the rest of China. The nudge is marginally more effective than the fine in the rest of China (t-test, $p = 0.06$), while the effectiveness of the two policy interventions is indistinguishable in Hubei (t-test, $p = 0.9$). Note that the difference in the effectiveness of the nudge in the Hubei province and outside of it is not statistically significant (t-test, $p = 0.7$).\\

\noindent \textbf{Response to social environment.} A theoretical analysis of the social distancing game assuming self-interested rational individuals predicts that the individual propensity to do social distancing should be higher in the homogenous social environment. This stems from the fact that the density of connections is higher than in the superspreader case (see SI). This may, however, differ behaviorally because the diffusion of COVID-19, especially on the outset, was driven by superspreading events \cite{wang2020inference, liu2020secondary, AWWLTCLC_2020}. Figure \ref{fig:2a} shows that the theoretical prediction is validated in the rest of China -- participants’ propensity to do distancing is 11.1 percentage points higher in the homogeneous compared to the superspreader environment (t-test, $p = 0.02$). In contrast, participants from Hubei do as much distancing in the superspreader environment as they do in the homogeneous one -- the difference is only 1.8 percentage points and it is insignificant (t-test, $p = 0.5$).

Figure \ref{fig:2b} delves deeper into the behavior in the superspreader environment between Hubei and rest of China participants. It reports average propensity to practice distancing split by place of residence, type of intervention, and position in the social environment. The left panel shows that Hubei-based participants in the superspreader position do significantly more social distancing relative to those in the homogeneous environment, while recipients do about as much as those in the homogeneous social structure. This is true for both the baseline part of the experiment (middle set of bars), and the two interventions (top and bottom sets of bars). The right panel shows that the behavior of the participants in the rest of China is quite different. Here, subjects in the supersreader position do as much social distancing as in the homogeneous environment particularly in the baseline part of the experiment and following a nudge intervention. In contrast, the recipients perform significantly less distancing relative to both the superspreader and the homogeneous environment throughout the  experiment. 

To confirm these observations, we repeat our core analysis, but instead of using a single dummy for a superspreader environment, we include one for each of the types of positions in this environment. For full table with the coefficients, consult the SI. The results of this exercise confirm our observations. In Hubei the superspreader has a 8.9 percentage points higher propensity to social distance compared to participants in the homogeneous environment (t-test, $p = 0.02$), and peripheral participants do as much distancing as participants in the homogeneous environment (t-test, $p = 0.3$). In contrast, in the rest of China, superspreader participants do as much distancing as participants in a homogeneous environment (t-test, $p = 0.6$), and peripheral participants have a 14.7 percentage points lower propensity to social distance compared to the ones in a homogeneous environment (t-test, $p = 0.002$).\\

\noindent \textbf{Risk aversion and social preferences.} An alternative explanation for our findings on the behavioral differences between participants from Hubei and the rest of China is that the two subject pools differ in terms of their general attitudes toward risky behavior. In fact, past research shows that natural disasters can sometimes lead to persistent increases in risk aversion \cite{CJK_2017}. As part of the experiment, we collect subjects’ attitudes to risk using an incentive-compatible ‘Bomb’ risk elicitation task (BRET) \cite{CF_2013}. The task amounts to deciding how many boxes to collect from a maximum of 100, with more boxes translating into potentially higher earnings, but also a higher risk of collecting a (hidden) bomb that destroys all earnings. Theory predicts that a risk-neutral subject collects 50 boxes with lower values indicating greater risk aversion. The average subject in our sample is moderately risk-averse with a BRET score of 42, which is consistent with previous findings in the literature \cite{CF_2013}. Additionally, as part of recruitment, we collect subjects’ self-reported attitudes to risk \cite{WBB_2002}.

There is no difference in general risk attitudes between the Hubei and rest of China participants according to either the BRET score (Mann-Whitney test, $p = 1.0$) or the self-reported risk index (Mann-Whitney test, $p = 0.2$). In other words, the differences between Hubei and rest of China participants seem to be confined to behaviors related to the COVID-19 pandemic, i.e. social distancing, rather than general behavior. Interestingly, Figure \ref{fig:2a} shows that the propensity to do social distancing is increasing with risk aversion (as captured by the BRET score) for participants in the rest of China (t-test, $p = 0.02$), but there is no significant association for Hubei-resident participants (t-test, $p = 0.6$). A potential explanation is that the harsh experience of the lockdown and/or pandemic in Hubei generates a widespread attitude toward social distancing that is independent of generalized risk preferences, while in the rest of China the propensity to practice social distancing is, as one would expect, increasing with risk aversion.

A second alternative explanation for our findings is that participants from Hubei have a less self-interested attitude compared to participants from other parts of China, and therefore they practice more social distancing to benefit others. The study of the effects of the 2004 tsunami in Thailand suggests that ‘shock’ events may lead to more prosocial behavior \cite{CJK_2017}. We collect subjects’ social preferences using an incentivized 6-item Social Value Orientation (SVO) task \cite{MAH_2011}. The underlying idea of the SVO framework is that people vary in terms of their motivations when evaluating different allocations of resources between themselves and others. In our sample, 51\% of subjects are individualists and 49\% are prosocials, and therefore we use a binary variable to capture social preferences.

There is no difference in social preferences between participants from Hubei and the rest of China according to the SVO score (Mann-Whitney test, $p = 0.5$). This indicates that the observed differences in behavior by Hubei participants are specific to social distancing attitudes, rather than general differences in preferences. In general, we would expect that participants with prosocial values are more likely to practice social distancing in the experiment compared to individualists because social distancing benefits others in their group. Figure \ref{fig:2a} shows that subjects from Hubei province with prosocial values are marginally more likely to do social distancing with an average difference of 7.7 percentage points (t-test, $p = 0.07$), while the difference is only 0.6 percentage points in the rest of China and is not significant (t-test, $p = 0.9$).\\

\noindent \textbf{Demographic characteristics.} When it comes to demographic characteristics, we find two heterogeneities between Hubei province and the rest of China (Figure \ref{fig:2a}). First, while more education is associated with significantly more distancing outside of Hubei province, the effect is not present within Hubei. In particular, an extra year of education is associated with 5.2 percentage points more distancing in the rest of China (t-test, $p = 0.0005$), while in Hubei the estimated size of the effect is -0.5 percentage point and is insignificant (t-test, $p = 0.8$). Second, subjects from Hubei province who were either employed or owned a business at the time of conducting the experiment did significantly more social distancing in the experiment, but the same is not true outside of it. The estimated effect in Hubei is 13.1 percentage points (t-test, $p = 0.02$), and 4.4 percentage points in the rest of China (t-test, $p = 0.5$). Note that the effects of age and gender are not significant in our experiment. This
may be partly explained by the fact that our sample is restricted to adults. In particular, existing research into adolescents suggests that younger people may differ in their attitudes to lockdowns and social distancing relative to adult population \cite{ren2021psychological}.

\section{Discussion}

The sudden outbreak of COVID-19 in Wuhan in early 2020 demanded a quick and decisive response from the government. To limit the spread of the virus, and potentially save tens of thousands of lives, the Chinese government implemented a very strict lockdown which affected most of Hubei province and lasted almost three months. Our experiment is the first to shed light on the possible medium-/long-term effects brought by the outbreak and the associated lockdown.

Our first result is that the level of social distancing is significantly higher in our experiment in Hubei province than outside of it. Both the distance of one's place of residence from the epicenter of the outbreak -- Wuhan -- and the strictness of government response to COVID-19 during the initial lockdown are robust predictors of individual propensity to social distance. Our experiment does not attempt to discriminate between the two plausible explanations -- severity of the outbreak and strictness of lockdown -- but it is clear that the overall experience of Hubei residents has created a behavioral difference that is persistent half a year after the end of the outbreak.

Our second result is that a soft intervention in the form of an informational video (a nudge), which highlights the harm caused to others by not practicing social distancing, is effective at increasing individual propensity to social distance throughout China. In contrast, a hard intervention in the form of a fine for non-compliance seems to work in Hubei province but not outside of it.

Our third result is that subjects in Hubei and the rest of China react differently to a superspreader type of social environment. Participants in the rest of China largely respond in line with theoretical predictions based on a standard game-theoretic framework, while participants from Hubei province violate these predictions. In particular, in a superspreader environment the recipients do not decrease their social distancing relative to the homogeneous environment, while superspreaders increase their propensity to social distance. This leads to a higher level of social distancing overall.

We find that the above behavioral heterogeneities cannot be explained by standard measures of general risk aversion or social preferences. The data suggests that these differences in behavior are specific to the context of social distancing and COVID-19, rather than general differences in preferences between Hubei residents and those from the rest of the country. Note that our list of controls is non-exhaustive, and other potential explanations may contribute to the observed differences in behaviour. For example, it has been shown that in China people with anxiety and depression are more willing to pay for a COVID-19 vaccine \cite{hao2021attitudes}.

Finally, our study highlights the important role that interactive web-based experiments can play in investigating people’s behavior, and how behavior can be affected by ‘shock’ events. Even though the decision situation faced by subjects in our experiment is artificial, we find clear and robust differences in behavior of subjects from the Hubei province and the rest of China, even after controlling for demographic characteristics and social preferences. 

\section{Methods}

Full details on methods, including theoretical framework, data collection and data analysis methods, together with a detailed description of the dataset are in the Supplementary Information (SI).\\

\noindent \textbf{Ethical approval.} This research received ethical approval for the use of human subjects from the Faculty of Economics Ethical Committee (University of Cambridge, ref.UCAM-FoE-20-02) and the Department of Psychology Ethics Committee (Tsinghua University, ref.THU202019). The experiment was performed in accordance with the relevant guidelines and regulations. Informed consent was obtained from all subjects before participation. \\

\noindent \textbf{Software.} The experiment was coded in oTree (v2.2.4) \cite{CSW_2016} with a server hosted on Heroku (www.heroku.com). \\

\noindent \textbf{Recruitment and sessions.} Subjects were recruited using the local survey company Wenjuan which is affiliated with Zhongyan Technology. During recruitment, we collected information on basic demographics, including gender, age, and place of residence. The experiment was conducted between October 3rd and November 14th 2020, and involved a total of 30 sessions with 1-5 groups each. The experiment took an average of 59 minutes (s.d. 20 minutes) to complete, and subjects earned an average of 17.7 yuan (s.d. 3.5 yuan). Subjects remained anonymous throughout both recruitment and experiment, and repeat participation was not allowed. \\

\noindent \textbf{Learning effects.} We identify significant learning effects in the early rounds of baseline and intervention parts of the experiment. In effect, subjects tend to converge to a particular stable strategy (e.g. always practice social distancing) after several rounds of the experiment. Therefore, in the analysis we use the last 10 rounds of the baseline and intervention parts of the experiment, but all results are robust to using all data. Full details on convergence analysis and relevant robustness checks are in the SI.\\

\noindent \textbf{Statistical analysis.} Since our dependent variable is binary, our analysis relies on a Linear Probability Model \cite{G_1964, A_1984}, but the estimates are robust to using a Logit or a Probit model instead (see SI). 

%% file: tables/main_text_table.tex
\begin{table}[h]
\caption{Main regression results for individual propensity to social distance}
\label{tab:1}
\resizebox{\textwidth}{!}{%
\begin{tabular}{lcccccl}
\toprule
\textbf{Dependent   variable:} & \multicolumn{6}{c}{social distancing   (binary)} \\ \midrule
\textbf{Model:} & \multicolumn{2}{c}{M1} & \multicolumn{2}{c}{M2} & \multicolumn{2}{c}{M3} \\ \midrule
\multicolumn{7}{l}{\textbf{Independent   variables:}} \\
Fine treatment & 0.0341* & (0.0195) & 0.0343* & (0.0195) & 0.0343* & (0.0195) \\
Nudge treatment & 0.0605*** & (0.0148) & 0.0603*** & (0.0148) & 0.0603*** & (0.0148) \\
Superspreader environment & -0.0491 & (0.0314) & -0.0494 & (0.0312) & -0.0481 & (0.0311) \\
Distance from Wuhan (100's km) & -0.0070** & (0.0031) &  &  &  &  \\
Hubei residence (1 = yes) &  &  & 0.0852*** & (0.0311) &  &  \\
OxCGRT index &  &  &  &  & 0.0076** & (0.0034) \\
Constant & 0.192 & (0.2080) & 0.101 & (0.2110) & -0.389 & (0.3660) \\
No of observations: & \multicolumn{2}{c}{8,280} & \multicolumn{2}{c}{8,280} & \multicolumn{2}{c}{8,280} \\
No of subjects: & \multicolumn{2}{c}{414} & \multicolumn{2}{c}{414} & \multicolumn{2}{c}{414} \\ 
\midrule
\multicolumn{7}{l}{\multirow{6}{*}{\parbox{18cm}{Notes: Standard errors  (reported in parentheses) are clustered at the group level. Significance level: *** $p<0.01$, ** $p<0.05$, * $p<0.1$. \textbf{(a)} All regressions use Linear Probability Model. \textbf{(b)} All models include the following controls: gender dummy (female = 1), age, years of education, employed or entrepreneur dummy (yes = 1), religious dummy (yes = 1), risk score (as captured by BRET), prosocial values (as captured by SVO; yes = 1). \textbf{(c)} To account of learning effects, we discard the first 10 rounds of baseline and intervention, and only consider 20 remaining rounds per subject. See Materials and Methods for details. \textbf{(d)} One subject did not complete the post experimental questionnaire and BRET.}}} \\
\multicolumn{7}{l}{} \\ 
\multicolumn{7}{l}{} \\ 
\multicolumn{7}{l}{} \\ 
\multicolumn{7}{l}{} \\ 
\multicolumn{7}{l}{} \\ 
\multicolumn{7}{l}{} \\  \bottomrule
\end{tabular}%
}
\end{table}

%% file: sections/theory.tex
In this section we present the theoretical model underpinning our experimental design. We also present our calibration of the model for the experiment, and list the key hypotheses that we test in Section \ref{sec:results}.\\

\noindent \textbf{The model.} A set of $N={1,2,\ldots,n}$ of risk-neutral agents are located on an unweighted and undirected network $\mathbf{G}$. $G_{ij}=G_{ji}=1$ when there is a link between agents $i$ and $j$, else $G_{ij}=G_{ji}=0$. Agents simultaneously decide whether to practice social distancing, at a private cost $c>0$. 

One agent is then chosen to be exposed to COVID-19 uniformly at random. In what follows, we refer to her as patient zero. If partient zero practices social distancing, she becomes infected with COVID-19 with probability $\gamma < 1$, otherwise, she is infected for sure. Any infected agent who is not practicing social distancing, including patient zero, can pass COVID-19 through contagion to her healthy neighbors who are not practicing social distancing with probability $\alpha \in [0, 1]$. Any agent who practices social distancing cannot pass the disease to others through contagion, or get infected through contagion herself.

Once contagion is over, payoffs are calculated. A healthy agent receives a benefit $b > c$, while an infected agent earns $0$ benefit. Additionally, any agent who decided to practice social distancing pays a cost $c>0$ regardless of her infection status.

Define the subset of agents who practice social distancing $S \subseteq N$, and the probability that an individual agent $i$ is infected $p_{i|S_i}$. Observe that if $i\in S$ then $p_{i|S_i}=\gamma / n$, since $i$ can only become infected if she is patient zero. Assuming a self-interested risk-neutral agent framework, the expected payoff to agent $i$ depends on her actions and the subset of agents who practice distancing. Specifically: 

\begin{align}
    \pi_{i} =
        \begin{cases}
        (1- \gamma/n)b-c,   & \text{if } i \in S \\
        (1 - p_{i|S_i})b,   & \text{otherwise.}
        \end{cases}
\end{align}

We use the following concepts in our analysis -- the socially optimal subset of agents who are practicing distancing, and the pure strategy Nash equilibrium subset. We define the socially optimal set subset of agents practicing distancing as one that maximizes the total expected payoff of the group. In cases where the expected payoff of agent $i$ is unchanged regardless of whether of not she is in the subset, we assume that she is. The pure strategy Nash equilibrium subset of agents who practice social distancing is such that every agent who belongs to the subset weakly prefers to be in the subset, whereas any agent not in the subset prefers to be outside of the subset. 

As discussed below, for some parameterizations of the model, the socially optimal subset of agents practicing distancing and the Nash equilibrium subset need not coincide. This then creates inefficiencies in the form of under- or over-provision of social distancing which is effectively a public good \citeplatex{mas1995microeconomic}. One way to correct the inefficiency is by introducing a fine for not practicing social distancing $f > 0$. The fine then adjusts the relative attractiveness of practicing social distancing by modifying expected payoff of agent $i$ such that:  

\begin{align}
    \pi_{i} =
        \begin{cases}
        (1- \gamma/n)b - c,   & \text{if } i \in S \\
        (1 - p_{i|S_i})b - f,    & \text{otherwise.}
        \end{cases}
\end{align}

Note that since the fine increases the costs of not practicing social distancing, we should expect it to weakly increase the amount of social distancing in a network regardless of its architecture. \\

\noindent \textbf{Parameterization.} To narrow down the scope of our theoretical analysis to results that are relevant for this experiment, we fix some parameters of the model as follows. First, we focus on setups with five agents -- i.e. $n= 5$. Next, we consider two network architectures. One is a complete network, where all nodes are connected to each other, the other is a star network where the central node -- the hub -- is connected to all other nodes, and no further links are present. To fix ideas, in what follows, we refer to these networks as homogeneous and superspreader environments. We refer to nodes in the homogeneous environment as `H', while those in the superspreader environment are `S' for the superspreader, and `R' -- for the recipient. Further, we set the (1) cost of practicing social distancing $c = 35$ points, (2) the benefit from being healthy $b = 100$ points, (3) the probability that patient zero who practices distancing is infected $\gamma = 0.5$, and (4) the probability with which an infected agent can pass COVID-19 to other agents not practicing distancing $\alpha = 0.65$. Finally, in some of our treatments we set the fine for not practicing social distancing $f = 15$ points (and $0$ in other treatments). In other treatments, we replace the fine by a nudge -- i.e. a 3-minute video which highlights the harm to others of not practicing distancing.

Consequently, we have a $2 \times 2$ full factorial design with two social environments -- homogeneous and superspreader, -- and two types of intervention -- fine and nudge. Note that all subjects in the experiment first play the game without an intervention, and then are subjected to one of the two interventions.\\

\noindent \textbf{Theoretical results.} Having set out the parameters, we can derive theoretical predictions of the model. Full set of proofs for our hypotheses is available from the authors on request.

\begin{theor}\label{theor:homogeneous_vs_superspreader}
   The average propensity to do social distancing is higher in the homogeneous environment compared to the superspreader.
\end{theor}

Given our parameterization, theoretical analysis of the above model predicts that in a pure strategy Nash equilibrium in a homogeneous environment 3 agents should practice social distancing. Note that since all agents are identical, there is no unique pure strategy Nash equilibrium subset. On the other hand, for the superspreader environment, the unique Nash equilibrium is such that only the superspreader practices distancing but not the recipients. It follows, that the average individual uptake of social distancing in the homogeneous environment is $0.6$, whereas in the superspreader environment it is $0.2$.

\begin{theor}\label{theor:social_optimality_vs_nash}
   There is under-provision of social distancing in the homogeneous environment relative to the social optimal but not in the superspreader environment.
\end{theor}

Theoretical analysis of the above model suggests that the socially optimal subset of agents who practice social distancing is any 4 of the 5 agents.  Given that the Nash equilibrium involves 3 agents practicing social distancing, the model predicts an under-provision of social distancing. Conversely, in the superspreader environment, the socially optimal subset of agents who practice distancing coincides with the Nash equilibrium subset -- i.e. only the superspreader practices distancing. Therefore, there should be no under- or over-provision of distancing in the superspreader environment. 

\begin{theor}\label{theor:fine_effect}
   A fine $f = 15$ points for not practicing social distancing increases the amount of social distancing. 
\end{theor}

We expect the fine to increase the amount of social distancing both in the homogeneous and superspreader environments. Note that the calibrated size of the fine is such that it should correct the inefficiency in the homogeneous environment, and have no effect or actually create an over-provision of social distancing in the superspreader environment. While the fine alters expected payoffs by making not practicing distancing more costly, the nudge highlighting the harm to others of not practicing distancing has no impact on the expected payoff of a self-interested rational agent. We therefore expect the nudge to have no effect on behavior regardless of the environment.

\begin{theor}\label{theor:nudge_effect}
   A nudge highlighting the harm to others of not practicing social distancing has no effect on agents' social distancing decisions. 
\end{theor}

Consequently, we expect the fine to be more effective than the nudge.

\begin{theor}\label{theor:fine_vs_nudge}
    A fine $f = 15$ points increases the amount of social distancing weakly more than the nudge.
\end{theor}

Finally, we hypothesize that real-world experience of a severe outbreak of COVID-19 may have an effect on behavior. Existing literature identifies behavioral effects of interventions, and generally `shock' events that persist well beyond the actual duration of the policies themselves \citeplatex{AJR_2005a, PLR_1993a}. We therefore think that the Hubei experience of the COVID-19 pandemic together including the strict lockdown that suddenly severely limited movement of 11 million citizens of Wuhan may have persistent effects on behavior of Hubei's residents, even after the lockdown was lifted on May 2nd 2020. In particular, we hypothesize that Hubei resident may exhibit greater risk aversion when it comes to contagious environments, and so practice more social distancing in our stylized game. We formulate the following research question.

\begin{quest}\label{quest:hubei}
   Does the experience of the outbreak of the COVID-19 pandemic and a total lockdown in Hubei province increase the amount of social distancing?
\end{quest}

To investigate this research question, we sample roughly half of our subject pool from the Hubei province, and the other half -- broadly from the rest of China. Further details are in Section \ref{sec:dataset}.

%% file: sections/methods_collection.tex
In this section we describe our data collection procedures. Section \ref{sec:recruitment} explains how we recruited subjects. Section \ref{sec:experiment} and Section \ref{sec:implementation} explain the workflow and implementation of the experiment.

\subsubsection{Recruitment} \label{sec:recruitment}

We recruited subjects using Wenjuan -- a local recruitment company.\footnote{Wenjuan is affiliated with Zhongyan Technology, see \url{https://www.wenjuan.com/}.} Subjects were sourced from 20 cities across 11 provinces to give us a diverse sample that is broadly representative of the urban population of China in terms of their geographical location and gender. Potential subjects were asked to fill in a quick survey and complete a qualifier quiz. 

The recruitment survey takes an average of 5 minutes to complete and pays a fixed reward of 5 yuan. As part of the survey, we collect information about participants' age, gender, province and city of residence, experience with decision-making experiments, and self-reported attitudes to risk \citeplatex{WBB_2002a}. Additionally, we inform participants about the upcoming interactive experiment and collect their consent for participation. Further, subjects are asked 2 (out of 4 randomly chosen) questions aimed at testing basic understanding of probability theory. Participants are given 3 attempts at the questions and must answer both correctly.

Participants who correctly answer the qualifying questions take part in a bonus task for a chance to win an amount in the 1.5-10 yuan range. The bonus task is the 6-item Social Value Orientation (SVO) task \citeplatex{MAH_2011a}. The task is aimed at eliciting subjects' social preferences, and classifies subjects as belonging to one of the four categories -- individualistic, competitive, prosocial, and altruistic. In practice, for each of the 6 decisions of the SVO task, participants are asked to choose between 9 different allocations of money between themselves and another anonymous person. These preferences are then used to determine subjects' types. For every 50 participants who completed the bonus task, we randomly selected 2 and implemented one randomly drawn decision of theirs. The instructions for the SVO scale along with screenshot of the interface from the recruitment survey are in Section \ref{sec:instructions_svo}. 

\subsubsection{Experiment} \label{sec:experiment}

Subject receives a link to our portal at the time advertised to her in an invitation. To join the session, she needs to click on the link and authenticate on our portal using a random ID issued to her during recruitment. As soon as she authenticates, she starts working through the instructions for the first part of the experiment (henceforth, Baseline). Instructions are followed by an understanding quiz, which the participant must pass to qualify for the experiment. The quiz has 3 question, and the participant must answer all of them correctly. Subject has 3 attempts at the quiz, and every unsuccessful attempt is followed by an explanation of correct answers. The instructions and the quiz take an average of 8.3 minutes to complete (s.d. 5.2 minutes). Throughout the instructions, subject is primed to think about COVID-19. Full instructions together with the quiz are in Section \ref{sec:instructions_main_experiment}. After passing the quiz, the participant joins a waiting room where she waits to be matched in a group with other subjects. Subject is compensated for waiting at a rate of 0.2 yuan for every 20 seconds of waiting up to a maximum of 5 yuan. Waiting time is capped at 10 minutes, and if the subject is not allocated into a group within 10 minutes, she receives compensation for reading the instructions, passing the quiz and waiting. Group allocation is randomized. Once a group is formed, participants proceed to Baseline where they play 20 rounds of the same game.

Each round of the game begins with subjects being randomly allocated to five positions in the environment -- either homogenous or superspreader -- which is constant throughout the experiment. Subjects learn their position in the environment and are asked to privately make their social distancing decisions at a cost $c = 35$ points. In each round, a subject has $80$ seconds to make their decisions, otherwise she receives a penalty of $50$ points. Failure to submit decisions in three consecutive rounds results in disqualification from the experiment without compensation. 

Once decisions are made, one subject is randomly selected to be patient zero. If patient zero practices distancing, she becomes infected with COVID-19 with probability 50\%, else she is infected for certain. If patient zero does not practice distancing, COVID-19 then spreads through the environment through contagion to other participants who chose not to practice distancing uniformly at random with probability 65\%. Note that those who practice distancing, cannot infect others or themselves become infected through contagion.

Once the contagious process is over, payoffs for the round are calculated. Healthy subjects earn $100$ points, while those infected earn zero. Further, all subjects who decided to practice distancing pay $35$ points regardless of their infection status. At the end of each round, subjects learn their outcomes for the round, and can also see their history of play for the last 5 rounds. Subjects have $20$ seconds to review this information. Note that subjects are not informed of the identity and/or decisions and outcomes of other members of their group at any point during or after the experiment. This is done to suppress effects of others' decisions and outcomes on individual choice.  

After the Baseline part of the experiment, participants proceed to the instructions for the Intervention, which depend on the treatment. In the fine treatment, subjects are explained that, in the next part of the experiment, they will receive a fine of $f = 15$ points every time they do not practice social distancing regardless of their infection status. In the nudge treatment, subjects must watch a 3-minute video which highlights the costs to others of not practicing social distancing.  Instructions for the intervention take on average 2.7 minutes (s.d. 1.7 minutes)\footnote{The nudge instructions take on average twice as long as the fine.}. Further details and instructions for the two types of intervention are in Section \ref{sec:instructions_main_experiment}. In both cases, the instructions are followed by a 1-question understanding quiz, asking subjects to identify the difference between Baseline and Intervention. The quiz serves essentially as an attention check, and participants have 3 attempts to answer the question. Once all members of a group pass the quiz, they proceed to Intervention. 

In Intervention, participants play further 20 rounds of the same game. The group composition and the social environment remain the same as in the Baseline. Further, for fine treatments, subjects receive a fine $f = 15$ points in every round where they decided not to practice distancing. For treatments with the nudge, the payment structure remains unchanged from the Baseline. 

Once subjects complete Intervention, the interactive part of the experiment is over. Subjects then proceed to the Post-experimental Questionnaire. Here, we ask a set of standard demographics questions. We repeat some of the questions from recruitment survey -- i.e. about gender and age -- to check answers for consistency.

Finally, a subject completes a Bomb Risk Elicitation Task (BRET) to elicit risk preferences \citeplatex{CF_2013a}. Subject is presented with 100 boxes arranged on a 10 $\times$ 10 matrix. One of these boxes contains a bomb but the location of the bomb is unknown. She is asked to choose how many boxes she want to collect. Boxes are collected from the top-right corner of the matrix, left to right, at a rate of one box per second. Participant must decide when to stop collecting boxes. After the boxes are collected, the contents of the boxes are revealed. If the bomb is collected, it explodes and reduces participant's earnings for BRET to zero. Otherwise, the participant earns 0.1 yuan for each collected box. Assuming a power utility function\footnote{Utility of payoff $x$ is defined as $u(x)=x^r$, where $r$ is the risk aversion coefficient.}, a risk-neutral participant opens 50 boxes. If a participant opens more than 50 boxes she is considered risk-seeking, and if she opens fewer than 50 boxes, she is considered risk-averse. Instructions for the BRET together with screenshots of the interface are in Section \ref{sec:instructions_bret}.

Upon completing BRET, subject is taken to the Payment Page, where she can see her earnings for the experiment. She can also scroll through her history of play in Baseline and Intervention. All participants receive a fixed fee of 5 yuan. Additionally, subjects earn a bonus for the interactive part of the experiment, BRET, and any waiting time if applicable. To reduce wealth effects \citeplatex{CGH_2016}, for the interactive part of the experiment subjects are paid for 4 randomly chosen rounds of Baseline and Intervention. Earnings are converted at a rate of 50 points per 1 yuan. 

On average, the experiment takes 59 minutes (s.d. 20 minutes) to complete and pays 17.7 yuan (s.d. 3.5 yuan) including a 5 yuan fixed fee. Upon completing the experiment, participants receive a code to submit to the survey company who then process payments.

\subsubsection{Implementation} \label{sec:implementation}

We conducted the experiment between October 3rd and November 14th 2020. The main experiment was programmed in o-Tree (v2.2.4; \url{www.otree.org})~\citeplatex{CSW_2016a} with a server deployed on Heroku (\url{www.heroku.com}). For the BRET task, we used a modified version of the implementation by Holzmeister and Pfurtschelle \citeplatex{HP_2016}.

We collected the data in a total of 30 sessions, each with 1-5 groups. Session assignment to treatment was randomized. A typical session had 25-60 places which were allocated on the first-come-first served basis. We accepted new subjects for 10 minutes since the start of the session, or until no more spaces were available.

Only those participants who correctly answer qualifying questions in the recruitment survey and give their consent for participation in the experiment were invited to the main experiment. Further, we kept track of IP addresses of subjects who have already completed the experiment, and excluded those with duplicate IP addresses from the list of invitees. Finally, as explained in Section \ref{sec:experiment}, subjects must pass two understanding quizzes during the main experiment to qualify for participation.

When running  pilots for this experiment, we discovered that a small proportion (4\%) of subjects suffered from random latency issues. Consequently, even though they were able to join the experiment and get allocated to a group, they were unable to play normally as the interface would not load on their side within the allocated amount of time. Extending the allocated time beyond 80 seconds per decision and 20 seconds per review proved to not improve performance. As a result and to avoid loosing too many groups, we decided to introduce `ghost' subjects, who could step in and take the place of those who dropped out. In practice this worked as follows.

We allocate subjects into groups of 6 rather than 5, and assign one subject to be a `ghost'. The main 5 subjects play the game just as described above. The experience for the `ghost' is very similar. In every round of the game, the `ghost' is randomly assigned to one position in the environment. Once everyone in the group makes their decisions, the outcome of the ghost is decided based on her actions and the actions of the subjects in the group who occupy the other four positions in the environment. This way, actions of the `ghost' subject do not affect the outcomes of the group as long as no dropout occurs, but the ghost receives the experience of being part of that group. If a dropout occurs, the `ghost' simply overtakes as part of the main group and the experiment proceeds as normal. The process is entirely seamless for all subjects. Consequently, we allow for up to 20\% change in the group composition and preserve the group. 

Of our 83 groups, 18 experienced a dropout overtaken by a ghost. 8 of those occurred in Baseline, 9 between Baseline and Intervention (i.e. due to a participant in the `main' group failing the understanding quiz), and 1 in Intervention. In our main analysis, we consider data from 415 subjects including those 18 `ghosts' but excluding data from the dropouts.  As a robustness check, we check that the results are not sensitive to alternative specifications, including using data from dropouts before they were disqualified and excluding data from groups with dropouts altogether.

%% file: sections/dataset.tex
Our dataset contains decision data of $415$ participants. Each subject participated in exactly one session. For all treatments, we collected data for at least $20$ groups of $5$ participants each, and roughly half of the groups were from the Hubei province.\footnote{We collected an 11th group from Hubei province with a fine intervention in the superspreader environment, and 2 further groups from the rest of China in the homogeneous environment with a fine intervention.} In each treatment, participants interacted for $40$ rounds ($20$ Baseline and $20$ Intervention), to a total of $16,600$ decisions. We also match the experimental data with data from the recruitment survey.

Apart from data on subjects' decisions in the experiment, we collect data on a set of variables, which can be broadly categorized as follows: demographics, preferences, and location-based controls. \Cref{tab:controls} presents summary statistics for some of these controls.\\

\input{tables/design_data.tex}

\noindent \textbf{Demographic controls.} All participants are resident in China, $47\%$ are female, and the mean age is $35$ years. An average subject in our sample has $18.7$ years of education.\footnote{We estimate years of education using subjects' highest qualification reported. We assume that all subjects took the standard number of years to complete each qualification, and undertook no education that did not lead to a qualification.} $76\%$ of the sample are either employed or entrepreneurs. $14\%$ are religious, with $7.5\%$ and $1\%$ respectively identifying Buddhism and Taoism as their religion, and $5.5\%$ reporting to practice some other religion.\\

\begin{figure}[ht!]
    \centering
    \includegraphics[width=\linewidth]{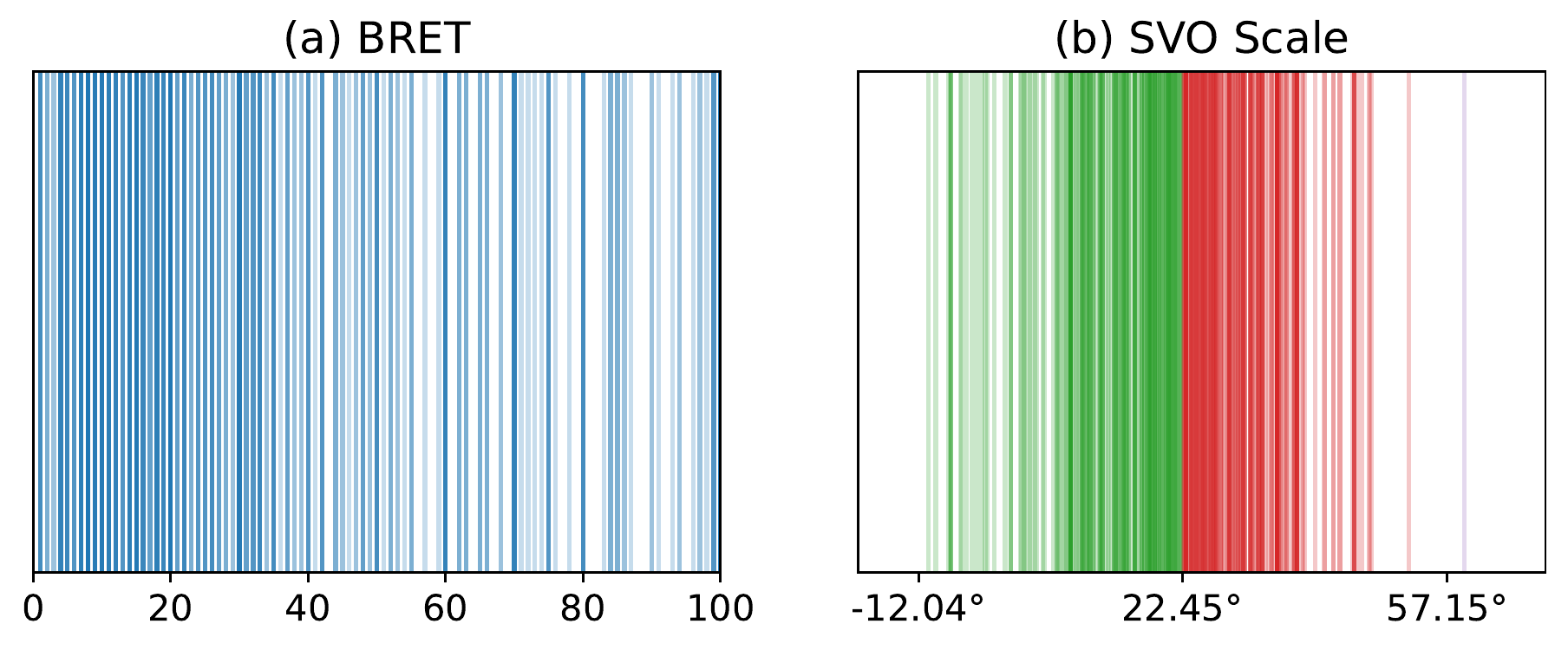}
    \caption{Bomb Risk Elicitation Task (BRET) and Social Value Orientation (SVO) Scale distributions for the sample. We draw a vertical line on the subplots for each subject whose score in BRET/SVO is of the corresponding value. More intense line color indicates that more subjects are concentrated at that value. For BRET, higher value corresponds to greater risk-seeking. For SVO the classification is as follows: angle $\geq$ 57.1$^{\circ}$ -- altruist; $\geq$ 22.45$^{\circ}$ and $<$57.15$^{\circ}$ -- prosocial; $<$22.45$^{\circ}$ and $\geq$ 12.04$^{\circ}$ -- individualist; $<$-12.04$^{\circ}$ -- competitive. Sample size is 414 (BRET), 415 (SVO).}
    \label{fig:distribution_bret_svo}
\end{figure}

\noindent \textbf{Preference controls.} Further, as explained in the previous sections, we collect information on subjects' social value (SVO) and risk (BRET) preferences. The distributions of both for our sample are summarized in \Cref{fig:distribution_bret_svo}. The average subject in the sample is moderately risk averse, with a BRET score of $42$ boxes. When it comes to social values, as captured by the SVO, $51\%$ of our subjects are classified as individualists. A further $49\%$ are prosocials. We have $1$ subject who classified as altruistic, and no subjects classified as competitive. For both BRET \citeplatex{CF_2013} and SVO \citeplatex{MAH_2011}, the resulting distributions are similar to the ones typically obtained in laboratory experiments.\\

\noindent \textbf{Location-based controls.} As part of recruitment, we collect data on participants province of residence, while in the experiment we collect subjects' IP-addresses, which we then use to back out the province. 
Figure 1 (see main text) shows counts of subjects per province (based on data from recruitment). The province recorded from recruitment and IP from the experiment differs for $37$ subjects in our sample. Note that this is not surprising, since, while IP data is 99\% accurate for country identification, it is only 50-80\% accurate for cities and regions. Further, the survey company responsible for recruitment has confirmed that some of those subjects were on a business trip when participating in the experiment, while others were using VPN. Excluding those subjects does not alter the distribution substantially. \\

%% file: tables/design_data.tex
\begin{table}[ht!]
\centering
\begin{threeparttable}
\caption{Summary statistics the main controls.}
\label{tab:controls}
\begin{tabular}{llccl}
\toprule
Variable & & 
\multicolumn{1}{c}{$\bar{X}$} & 
\multicolumn{1}{c}{s.d.} & 
Comments \\ \midrule
\multicolumn{5}{l}{\underline{Demographic controls}} \\
Age & &  $35.13$ & $10.23$ & measured in years  \\
Gender & & $0.47$ & $0.50$ & female = 1 \\
Education & & $18.7$ & $1.48$ & measured in years \\
Employed & & $0.76$ & $0.43$ & yes = 1 \\
Religious & & $0.14$ & $0.35$ & yes = 1 \\
\multicolumn{5}{l}{\underline{Preference controls}} \\
BRET score & & $41.76$ & $33.14$ & $\in [0,100]$ \\
SVO type & & $0.49$ & $0.50$ & prosocial = 1 \\ 
\multicolumn{5}{l}{\underline{Residence controls}} \\
Hubei & & $0.49$ & $0.50$ & resides in Hubei = 1 \\ 
Distance from Wuhan & & $4.92$ & $4.50$ & in 100's of kilometers \\ 
\bottomrule
\end{tabular}
\begin{tablenotes}
      \item Sample size is 414, because 1 subject did not complete post-experimental questionnaire and BRET; $\bar{X}$ -- mean value, or proportion in case of binary variables; s.d. -- standard deviation. 
    \end{tablenotes}
\end{threeparttable}
\end{table}

%% file: sections/methods_analysis.tex
\subsubsection{Convergence}\label{sec:analysis_convergence}

\begin{figure}[ht!]
    \centering
    \includegraphics[width=\textwidth]{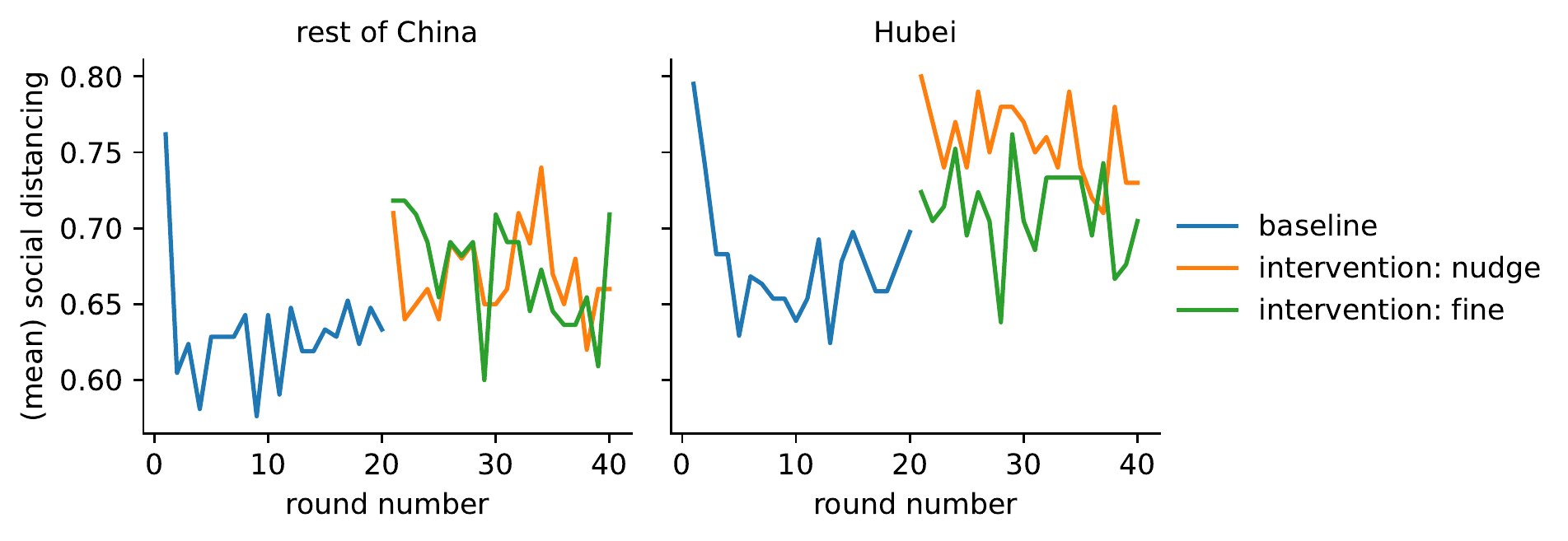}
    \caption{Mean distancing levels in Baseline and Intervention, separately for Hubei province and the rest of China.}
    \label{fig:mean_distancing}
\end{figure}

\Cref{fig:mean_distancing} presents the evolution of average propensity to practice distancing at the individual level, separately for Hubei province and the rest of China, in both Baseline and Intervention (separately for fine and nudge). 

Experiments on public good games typically exhibit a decline in contributions to the public good for several rounds of the experiment before contributions converge to a stable level \citeplatex{ledyard1994public}. Similarly, in our experiment, subjects exhibit a similar pattern in both Baseline and Intervention as shown in \Cref{fig:mean_distancing}. Since we are interested in the limiting outcomes of this convergence behavior, the first step of our analysis is to determine the cut-off round, after which the majority of subjects converge to a stable strategy. 
We define individual convergence as follows:

\begin{definition}
    A participant converges to a strategy $s$ by round $n$ if (i) she used this strategy for the last $k$ rounds (including $n$), and (ii) in all subsequent rounds $[n+1, 20]$ the number of consecutive deviations from the chosen strategy does not exceed $a$.
\end{definition}

We consider three types of convergence strategies. In both homogeneous and superspreader environments, we look at the strategy where the subject always chooses the same action. For the superspreader environment, we also consider two extra strategies. In one strategy the participant always chooses the same action when she is in the superspreader position and the complement action when she is a recipient. In the other strategy, she always chooses the same action when she is the superspreader and alternates between the two actions when a recipient.

We set $k = 4$ and $a=2$. Consequently, to be considered converged at a round, a subject must choose the same strategy for the last four rounds, and not deviate from that strategy for more that two consecutive rounds. Moreover, the earliest a subject can be considered to converge to a stable strategy is by round 4. 

\input{tables/analysis_convergence.tex}

\Cref{tab:convergence} summarizes our convergence analysis using the above definition. From the table, we can see that by round 11 at least 77\% of subjects in Baseline and 86.3\% in Intervention converge to a stable strategy. In fact, when disaggregated by treatment, the lowest and highest convergence rates are 70.7\% (star in Baseline) and 87.4\% (fine in Intervention). Further analysis in Section \ref{sec:robustness} shows that our convergence analysis is robust top using a more/less conservative definition of convergence.

\subsubsection{Parametric analysis}

To analyze determinants of social distancing behavior in our experiment, we employ the following econometric framework. We assume that individual social distancing decision -- a binary variable -- is a linear function of our treatment variables and a range of controls. Consequently, we have the following linear probability model (LPM):

\begin{align}
    y_{it} = x_{it} \beta + v_{i} + \epsilon_{it},
\end{align}
 
 where 

\begin{align}
    \text{Pr}(y_{it} = 1|x_{it}) = x_{it} \beta + v_{i}.
\end{align}

The model assumes that the subject-specific random effect $v_{i}$ is normally distributed, and error term $\epsilon_{it}$ is normally distributed and clustered at the group level.

The choice of a LPM over non-linear models is motivated by easy interpretation of coefficients, especially when it comes to interaction effects. However, as shown in Section \ref{sec:robustness}, our results are not sensitive to a choice of model, as the estimated average marginal effects for the random effects Logit and Probit models are very similar.

Since the majority of subjects in our dataset exhibit clear convergence in behavior, our econometric analysis utilizes data from rounds 11-20 of both Baseline and Intervention -- i.e. once participants behavior has converged to a stable strategy. In Section \ref{sec:robustness} we show that our key results are robust to using all of the data.

We perform econometric analysis in several stages, by consecutively adding controls which may plausibly explain social distancing decisions. To keep this section concise, full details of all our specifications are in Section \ref{sec:results}.

%% file: tables/analysis_convergence.tex
\begin{table}[h]
\centering
\begin{threeparttable}
\caption{Individual convergence analysis.}
\label{tab:convergence}
\begin{tabular}{cccccc}
\toprule
\multirow{2}{2cm}{Network} &
\multirow{2}{2.5cm}{\begin{tabular}[c]{@{}c@{}}Type of \\ intervention\end{tabular}} &
\multirow{2}{2cm}{Population} & \multirow{2}{1cm}{n} & \multirow{2}{3cm}{\begin{tabular}[c]{@{}c@{}}$>70\%$ converged\\ by round ...\end{tabular}} & \multirow{2}{3cm}{\begin{tabular}[c]{@{}c@{}}...$\%$ converged \\ by round 11\end{tabular}} \\
&  &  & \\
\midrule
\multicolumn{5}{l}{\textbf{Baseline}} \\ \hline
all & all & all & $415$ & $8$ & $77.8$ \\ \hline
complete & all & all & $210$ & $7$ & $83.8$ \\ 
star & all & all & $205$ & $10$ & $70.7$ \\ \hline
all & all & Hubei & $205$ & $8$ & $76.6$ \\ 
all & all & non-Hubei & $210$ & $9$ & $79.0$ \\ \hline
\multicolumn{5}{l}{\textbf{Intervention}} \\ \hline
all & all & all & $415$ & $6$ & $86.3$ \\ \hline
complete & all & all & $210$ & $5$ & $86.7$ \\ 
star & all & all & $205$ & $6$ & $85.9$ \\ \hline
all & all & Hubei & $205$ & $6$ & $73.2$ \\ 
all & all & non-Hubei & $210$ & $6$ & $83.8$ \\ \hline
all & fine & all & $215$ & $5$ & $87.4$ \\ 
all & nudge & all & $200$ & $6$ & $85$ \\
\bottomrule
\end{tabular}
\end{threeparttable}
\end{table}

%% file: sections/results.tex
\subsubsection{Overall effects}

In our shortest specification (F1 of \Cref{tab:main_regs}) we add dummies for experimental treatments: fine and nudge interventions, and the superspreader environment. As we randomly assign participants to these treatments, we are confident that their effects are causal. 

Next, we add controls for social demographics and risk preferences and social preferences (F2), and a variable that captures the distance of one's place of residence from Wuhan in 100's of kilometers (F3). 

The next specification (F4) is an auxilary one. Here, we replace distance from Wuhan in M1 with a dummy variable equal to one if the participant is from Hubei province and zero otherwise.

Further, using the data from the Oxford COVID-19 Government Response Tracker (OxCGRT), we construct an index of average overall government response for the period of 23rd Jan-2nd May 2020 \citeplatex{H_2020}. This results in a single value of an index for each of the provinces in our sample. The correlation between distance from Wuhan and this index for the 20 cities in our sample is -0.6683 (t-test, $p = 0.0001$). Our next specification (F5) replaces distance from Wuhan with this index. Note that specifications F3-F5 are the full versions of the reduced M1-M3 reported in the main text.

Finally, to better understand the impact of the superspreader environment, in F6 we replace the superspreader environment dummy with two dummies -- one for being in the superspreader position and another one for being the recipient.

Below we briefly summarize the main results of this exercise. We refer to effects that are significant at $1\%$ level, as being very significant, at $5\%$ level -- as significant, and $10\%$ level -- as marginally significant.\\

\input{tables/main_regressions}

\noindent \textbf{Interventions.} Fine for not practicing social distancing appears to work in China, but its overall effect in the experiment is about 3.4 percentage points and is marginally significant in all models. Conversely, the effect of the nudge is almost twice in magnitude (6.0 percentage points) and is very significant in all specifications.\\

\noindent \textbf{Environment.} Overall, the effect of the superspreader environment is not significant in F1-F5. However, unpacking the superspreader environment into superspreaders and recipients (F6), we can see that superspreaders practice 7.2 percentage points more social distancing than subjects in the homogeneous setting and the effect is significant. Conversely, recipients practice 8.0 percentage points less distancing than subjects in the homogeneous setting, again, with the effect being significant.\\

\noindent \textbf{Demographics.} According to \Cref{tab:main_regs}, an extra year of education is associated with 2.4 percentage points more social distancing. Further, employed subjects or those who run their own business, practice 10.4 percentage points more social distancing. Both effects are significant.\\ 

\noindent \textbf{Preferences.} Overall, we do not find evidence that subjects' risk preferences and social values affect their social distancing decisions.\\

\noindent \textbf{Distance from Wuhan.} We find that subjects who reside further away from Wuhan practice less social distancing in the experiment (F3). The estimated size of the effect is 7.0 percentage points for 1,000 km and is significant. We also find that residing in Hubei province results in 8.5 percentage points higher propensity to practice social distancing compared to the rest of China (F4).\\

\noindent \textbf{Strictness of government response.} Harsher COVID-19 related measures undertaken by the government during the initial lockdown are associated with more social distancing. Specifically, an extra 10 points on the average OxCGRT index contributes an increase of 7.6 percentage points in the individual propensity to social distance (F5). 

\subsubsection{Heterogeneities}

To better understand the driving forces behind the differences in the observed propensity to social distancing in Hubei province and outside of it, we perform the following analysis. We interact every variable in specification \Cref{tab:main_regs} F1 with the Hubei dummy and calculate the average marginal effects of each variable separately for Hubei and the rest of China. The resulting output is in \Cref{tab:interactions}.\\

\noindent \textbf{Interventions.} We find that fine for not practicing social distancing is effective in Hubei province but not outside of it. On the other hand, the nudge appears to be effective throughout China.\\

\noindent \textbf{Environment.} Theory predicts that there should be less social distancing in the superspreader environment, trivially due to the network of interactions being less dense. Subjects from Hubei province are, however, not responsive to the superspreader environment -- the amount of social distancing is not significantly different from that observed in the homogeneous environment. Outside of Hubei, however, subjects do less social distancing in the superspreader environment, and the effect is significant.\\

\noindent \textbf{Demographics.} Outside of Hubei province, more years of education is associated with more social distancing, and the effect is very significant. The same effect is not observed in Hubei province. On the other hand, subjects from Hubei who were employed or had their own business during the experiment, do significantly more social distancing, but the effect is not observed in the rest of China.\\ 

\noindent \textbf{Preferences.} We find that risk-seeking is associated with less social distancing outside of Hubei province with the effect being significant. Within Hubei province, the effect is not observable. Further, having prosocial values increases individual propensity to practice social distancing for subjects from the Hubei province but not the rest of China, with the effect being significant.\\

\input{tables/interactions_regressions}

%% file: tables/main_regressions.tex
\begin{table}[ht!]
\centering
\caption{Main regression results}
\label{tab:main_regs}
\resizebox{\textwidth}{!}{%
\begin{tabular}{@{}lcccccc@{}}
\toprule
\multicolumn{1}{r}{\textbf{Dependent variable:}} & \multicolumn{6}{c}{social distancing (1 = yes)} \\ \cmidrule{2-7}
\multicolumn{1}{r}{\textbf{Model$^a$}} & F1 & F2 & F3 & F4 & F5 & F6 \\
 \midrule
\multicolumn{7}{l}{\textbf{Independent variables:}} \\
Fine treatment & 0.0322* & 0.0341* & 0.0341* & 0.0343* & 0.0343* & 0.0343* \\
 & (0.0192) & (0.0196) & (0.0195) & (0.0195) & (0.0195) & (0.0195) \\
Nudge treatment & 0.0608*** & 0.0606*** & 0.0605*** & 0.0603*** & 0.0603*** & 0.0603*** \\
 & (0.0148) & (0.0148) & (0.0148) & (0.0148) & (0.0148) & (0.0149) \\
Superspreader environment & -0.0406 & -0.0440 & -0.0491 & -0.0494 & -0.0481 &  \\
 & (0.0322) & (0.0323) & (0.0314) & (0.0312) & (0.0311) &  \\
Superspreader &  &  &  &  &  & 0.0718** \\
 &  &  &  &  &  & (0.0317) \\
Recipient &  &  &  &  &  & -0.0796** \\
 &  &  &  &  &  & (0.0328) \\
Gender (1 = female) &  & 0.0126 & 0.0110 & 0.0060 & 0.0086 & 0.0120 \\
 &  & (0.0312) & (0.0311) & (0.0312) & (0.0311) & (0.0310) \\
Age &  & 0.0008 & 0.0014 & 0.0018 & 0.0019 & 0.0014 \\
 &  & (0.0016) & (0.0017) & (0.0017) & (0.0018) & (0.0017) \\
Years of education &  & 0.0233** & 0.0241** & 0.0243** & 0.0241** & 0.0246** \\
 &  & (0.0120) & (0.0119) & (0.0116) & (0.0117) & (0.0119) \\
Employed or entrepreneur (1 = yes) &  & 0.0920** & 0.1040** & 0.1050** & 0.1030** & 0.1030** \\
 &  & (0.0435) & (0.0430) & (0.0430) & (0.0427) & (0.0431) \\
Religious (1 = yes) &  & 0.0328 & 0.0344 & 0.0349 & 0.0308 & 0.0343 \\
 &  & (0.0473) & (0.0479) & (0.0476) & (0.0471) & (0.0479) \\
Risk score &  & -0.0003 & -0.0004 & -0.0004 & -0.0005 & -0.0004 \\
 &  & (0.0005) & (0.0005) & (0.0005) & (0.0005) & (0.0005) \\
Prosocial values (1 = yes) &  & 0.0295 & 0.0287 & 0.0286 & 0.0302 & 0.0279 \\
 &  & (0.0293) & (0.0289) & (0.0287) & (0.0287) & (0.0290) \\
Distance from Wuhan (100's km) &  &  & -0.0070** &  &  & -0.0071** \\
 &  &  & (0.0031) &  &  & (0.0031) \\
Hubei residence (1 = yes) &  &  &  & 0.0852*** &  &  \\
 &  &  &  & (0.0311) &  &  \\
OxCGRT index &  &  &  &  & 0.0076** &  \\
 &  &  &  &  & (0.0034) &  \\
Constant & 0.670*** & 0.196 & 0.192 & 0.101 & -0.389 & 0.184 \\
 & (0.0208) & (0.2080) & (0.2080) & (0.2110) & (0.3660) & (0.2080) \\ \midrule
No of observations: & 8,300 & 8,280 & 8,280 & 8,280 & 8,280 & 8,280 \\
No of subjects$^b$: & 415 & 414 & 414 & 414 & 414 & 414\\ \midrule
\multicolumn{7}{l}{\multirow{3}{*}{\parbox{19cm}{Notes: Standard errors (reported in parentheses) are clustered at the group level. *** $p<0.01$, ** $p<0.05$, * $p<0.1$. \textbf{(a)} All regressions use Linear Probability Model. \textbf{(b)} One subject did not complete the post experimental questionnaire and BRET.}}} \\
\multicolumn{7}{l}{}\\
\multicolumn{7}{l}{}\\
\bottomrule
\end{tabular}%

}
\end{table}

%% file: tables/interactions_regressions.tex
\begin{table}[h]
\centering
\caption{Average marginal effects separately for the rest of China and Hubei province}
\label{tab:interactions}
\begin{tabular}{@{}llllll@{}}
\toprule
\textbf{Dependent variable:} & \multicolumn{5}{c}{social distancing (1 = yes)} \\
\cmidrule{2-6}
 & \multicolumn{2}{c}{rest of China} & \multicolumn{1}{c}{} & \multicolumn{2}{c}{Hubei province} \\
 \midrule
\multicolumn{5}{l}{\textbf{Independent variables:}} &  \\
Fine treatment & 0.0090 & (0.0248) &  & 0.0602** & (0.0290) \\
Nudge treatment & 0.0671*** & (0.0189) &  & 0.0544** & (0.0226) \\
Superspreader environment & -0.1109** & (0.0459) &  & -0.0184 & (0.0393) \\
Gender (1 = female) & -0.0578 & (0.0416) &  & 0.0593 & (0.0466) \\
Age & 0.0020 & (0.0021) &  & 0.0008 & (0.0027) \\
Years of education & 0.0459*** & (0.0137) &  & -0.0041 & (0.0168) \\
Employed or entrepreneur (1 = yes) & 0.0450 & (0.0719) &  & 0.1313** & (0.0582) \\
Religious (1 = yes) & 0.0919 & (0.0598) &  & -0.0055 & (0.0682) \\
Risk score & -0.0015** & (0.0007) &  & 0.0003 & (0.0007) \\
Prosocial values (1 = yes) & 0.0087 & (0.0390) &  & 0.0775* & (0.0431) \\
\midrule
No of observations: & \multicolumn{5}{c}{8,280} \\
No of subjects: & \multicolumn{5}{c}{414} \\
\midrule
\multicolumn{6}{l}{\multirow{3}{*}{\parbox{16cm}{Notes: Standard errors (reported in parentheses) are clustered at the group level. *** $p<0.01$, ** $p<0.05$, * $p<0.1$. The regression is a Linear Probability Model. One subject did not complete the post experimental questionnaire and BRET.}}} \\
\multicolumn{6}{l}{} \\
\multicolumn{6}{l}{} \\ 
\bottomrule
\end{tabular}%

\end{table}

%% file: sections/further_analysis.tex
This section contains a selection of robustness checks. Section \ref{sec:robustness_main} reports robustness checks for our main results, while Section \ref{sec:robustness_convergence} contains further checks for our convergence analysis.

\subsubsection{Main results} \label{sec:robustness_main}

This section present 6 robustness checks on our main results, presented in \Cref{tab:main_regs_robustness}. Each is discussed in turn.

In R1 of \Cref{tab:main_regs_robustness} we re-estimate F4 from \Cref{tab:main_regs} using all data. From the table we can see that none of the estimates of the key parameters of interest change materially. Re-estimating any other model from \Cref{tab:main_regs} using all data results in the qualitatively similar observation. This suggests that our results are not sensitive to discarding data pre-convergence.

Next, R2 reports marginal effects of each of the variables from specification in F4 using a logistic regression in place of a linear probability model. Similarly, R3 reports marginal effects from a probit regression applied to the same specification. As we can see, the point estimates of all parameters remain practically unchanged. These specifications suggests that our results are not driven by the choice of model.

Recall that 18 of our groups experienced one subject dropping out midway through the experiment. As explained in Section \ref{sec:methods_collection}, we replaced these dropouts by a `ghost' subject. In R4 we re-estimate F4 but discard data from all groups where a `ghost' was activated due to a subject dropping out. Again, none of the estimates change materially, meaning that our experimental procedure of replacing dropouts does not drive our results.

Next, in our main analysis we use self-reported residence data to determine subjects' place of residence. It is possible, that this data is not very reliable. To investigate this possibility, we re-estimate specification in F3 now using location data as determined by subjects' IP-addresses when completing the experiment. Note that for 4 subjects in our sample the IP-address identified country as being other than China (e.g. Singapore of Thailand). Based on the information that we have, these subjects were using VPN when completing the study. Consequently, we exclude them from the analysis when estimating R5. Again, the obtained coefficients are not materially different from those in F3.

Finally, when looking at the association between OxCGRT index and social distancing behavior in the experiment we use data for the period 23rd Jan 2020 - 02nd May 2020. As a robustness check, in R6 we estimate the same index for the period 23rd Jan 2020 - 02nd May 2021 and use it as a control. This index is completely insignificant.

\input{tables/robustness_checks}

\subsubsection{Convergence} \label{sec:robustness_convergence}

\begin{figure}[ht!]
    \centering
    \includegraphics[width=\textwidth]{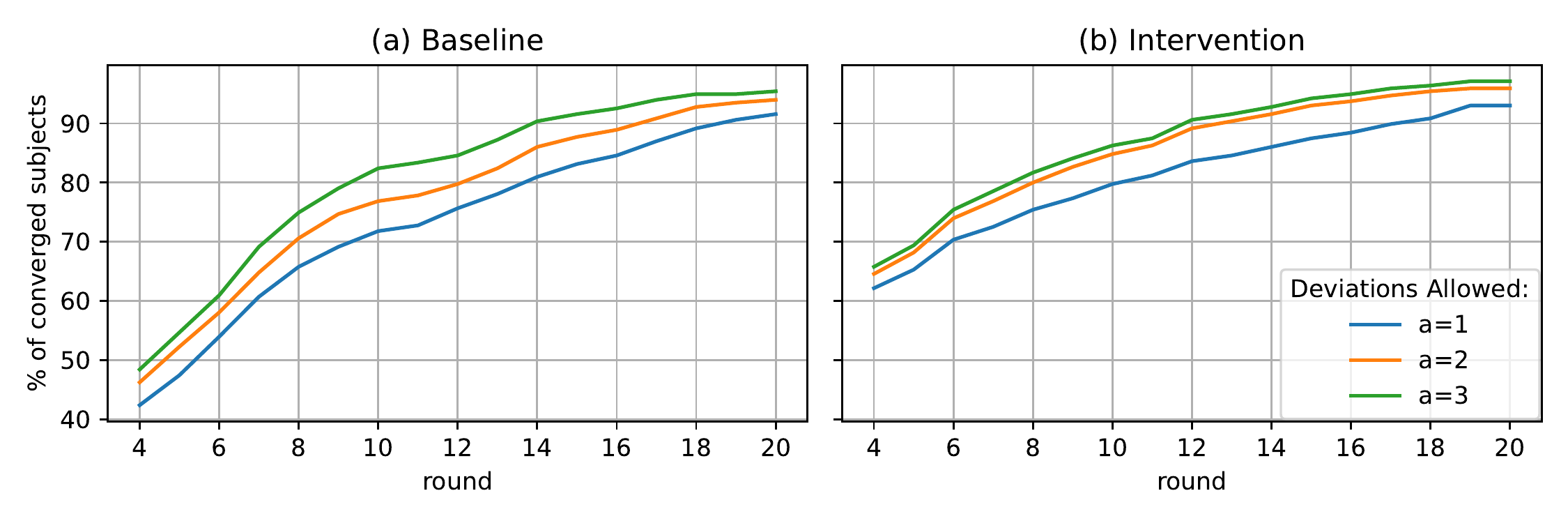}
    \caption{Evolution of the share of converged participants throughout the experiment separately for Baseline and Intervention, $a \in [1,3]$. Note: we exclude the first 3 rounds in both parts since $k=4$.}
    \label{fig:convergence}
\end{figure}

Recall that in Section \ref{sec:analysis_convergence} we define a subject to be converged to a stable strategy by a particular round if she followed this strategy for the last four rounds and in subsequent rounds does not deviate from this strategy for more than two consecutive rounds. As a robustness check, we consider allowing for one and three consecutive deviations respectively. \Cref{fig:convergence} plots the share of converged participants for each round separately by parts when allowing for 1-3 consecutive deviations ($a \in [1,3]$). We can see that the share of converged subjects does not change much when we allow for a more/less conservative definition. In particular, with $a=1$ the share of subjects who converge by round 11 in Baseline drops to $72.7\%$ while in Intervention it reaches $81.2\%$. With $a=3$ the share of subjects who converge by round 11 in Baseline and Intervention stands at $83.4\%$ and $87.5\%$ respectively.

As an extra robustness check for our convergence analysis, we recalculate convergence statistics for our sample, excluding the data from our 18 `ghost' subjects. The results are summarized in \Cref{tab:robustness_convergence}. We can see, that, with `ghosts' excluded, the share of converged subjects rises both in Baseline (from $77.8\%$ to $87.4\%$ ) and Intervention (from $86.3\%$ to $89.9\%$).

The above analysis suggests that it is reasonable to claim that the absolute majority of subjects converge to a particular strategy by round 11 in both parts of the experiment.

\input{tables/robustness_convergence}

%% file: tables/robustness_checks.tex
\begin{table}[ht!]
\centering
\caption{Main regression robustness checks}
\label{tab:main_regs_robustness}
\resizebox{\textwidth}{!}{%
\begin{tabular}{@{}lcccccc@{}}
\toprule
\multicolumn{1}{r}{\textbf{Dependent variable:}} & \multicolumn{6}{c}{social distancing (1 = yes)} \\ \cmidrule{2-7}
\multicolumn{1}{r}{\textbf{Model$^a$:}} & R1 & R2 & R3 & R4 & R5 & R6 \\
 \midrule
\multicolumn{7}{l}{\textbf{Independent variables:}} \\
Fine treatment & 0.0344** & 0.0343* & 0.0324* & 0.0452** & 0.0339* & 0.0340* \\
 & (0.0158) & (0.0193) & (0.0189) & (0.0218) & (0.0196) & (0.0196) \\
Nudge treatment & 0.0658*** & 0.0546*** & 0.0533*** & 0.0729*** & 0.0547*** & 0.0606*** \\
 & (0.0134) & (0.0136) & (0.0137) & (0.0172) & (0.0152) & (0.0148) \\
Superspreader environment & -0.0493* & -0.0455 & -0.0450 & -0.0432 & -0.0568* & -0.0437 \\
 & (0.0293) & (0.0298) & (0.0307) & (0.0353) & (0.0315) & (0.0323) \\
Gender (1 = female) & -0.00103 & -0.00730 & -0.00668 & -0.00108 & 0.0134 & 0.00975 \\
 & (0.0295) & (0.0305) & (0.0315) & (0.0371) & (0.0309) & (0.0312) \\
Age & 0.00147 & 0.00197 & 0.00201 & 0.00147 & 0.00130 & 0.000791 \\
 & (0.00159) & (0.00161) & (0.00165) & (0.00190) & (0.00169) & (0.00159) \\
Years of education & 0.0249** & 0.0217* & 0.0212* & 0.0154 & 0.0247** & 0.0231* \\
 & (0.0107) & (0.0112) & (0.0118) & (0.0120) & (0.0120) & (0.0119) \\
Employed or entrepreneur (1 = yes) & 0.0995*** & 0.0970** & 0.0957** & 0.133*** & 0.107** & 0.0925** \\
 & (0.0386) & (0.0400) & (0.0412) & (0.0459) & (0.0428) & (0.0436) \\
Religious (1 = yes) & 0.0384 & 0.0347 & 0.0337 & 0.0849* & 0.0430 & 0.0345 \\
 & (0.0446) & (0.0474) & (0.0490) & (0.0491) & (0.0477) & (0.0476) \\
Risk score & -0.000314 & -0.000502 & -0.000501 & -0.000387 & -0.000508 & -0.000324 \\
 & (0.000478) & (0.000497) & (0.000514) & (0.000560) & (0.000500) & (0.000513) \\
Prosocial values (1 = yes) & 0.0281 & 0.0336 & 0.0341 & 0.0292 & 0.0340 & 0.0290 \\
 & (0.0269) & (0.0270) & (0.0279) & (0.0317) & (0.0290) & (0.0293) \\
Hubei residence (1 = yes) & 0.0854*** & 0.0789*** & 0.0783*** & 0.0680* &  &  \\
 & (0.0285) & (0.0293) & (0.0302) & (0.0352) &  &  \\
Distance   from Wuhan (100's km) &  &  & & & -0.00675** &  \\
 &  &  &  &  & (0.00322) &  \\
OxCGRT index for 2021 &  &  &  &  &  & -0.00283 \\
 &  &  &  &  &  & (0.00362) \\ 
Constant & 0.109 & - & - & 0.243 & 0.187 & 0.349 \\
 & (0.196) & - & - & (0.226) & (0.209) & (0.300) \\ \midrule
No of observations: & 16,560 & 8,280 & 8,280 & 6,480 & 8,200 & 8,280 \\
No of subjects$^b$: & 414 & 414 & 414 & 324 & 410 & 414\\ \midrule
\multicolumn{7}{l}{\multirow{3}{*}{\parbox{19cm}{Notes: Standard errors (reported in parentheses) are clustered at the group level. *** $p<0.01$, ** $p<0.05$, * $p<0.1$. \textbf{(a)} R2 is a Logistic regression and R3 is a probit regression. Other models use a Linear Probability Model. \textbf{(b)} One subject did not complete the post experimental questionnaire and BRET.}}} \\
\multicolumn{7}{l}{}\\
\multicolumn{7}{l}{}\\
\bottomrule
\end{tabular}%

}
\end{table}

%% file: tables/robustness_convergence.tex
\begin{table}[ht!]
\centering
\begin{threeparttable}
\caption{Individual convergence analysis: robustness check}
\label{tab:robustness_convergence}
\begin{tabular}{cccccc}
\toprule
\multirow{2}{2cm}{Network} &
\multirow{2}{2.5cm}{\begin{tabular}[c]{@{}c@{}}Type of \\ intervention\end{tabular}} &
\multirow{2}{2cm}{Population} & \multirow{2}{1cm}{n} & \multirow{2}{3cm}{\begin{tabular}[c]{@{}c@{}}$>70\%$ converged\\ by round ...\end{tabular}} & \multirow{2}{3cm}{\begin{tabular}[c]{@{}c@{}}...$\%$ converged \\ by round 11\end{tabular}} \\
&  &  & \\
\midrule
\multicolumn{5}{l}{\textbf{Baseline}} \\ \hline
all & all & all & $397$ & $7$ & $87.4$ \\ \hline
complete & all & all & $202$ & $7$ & $83.7$ \\ 
star & all & all & $195$ & $6$ & $91.3$ \\ \hline
all & all & Hubei & $198$ & $7$ & $86.4$ \\ 
all & all & non-Hubei & $199$ & $7$ & $88.4$ \\ \hline
\multicolumn{5}{l}{\textbf{Intervention}} \\ \hline
all & all & all & $397$ & $5$ & $89.9$ \\ \hline
complete & all & all & $202$ & $5$ & $87.6$ \\ 
star & all & all & $195$ & $5$ & $92.3$ \\ \hline
all & all & Hubei & $198$ & $5$ & $90.4$ \\ 
all & all & non-Hubei &  $199$ & $5$ & $89.4$ \\ \hline
all & fine & all & $207$ & $4$ & $90.3$ \\ 
all & nudge & all & $190$ & $5$ & $89.5$ \\
\bottomrule
\end{tabular}
\end{threeparttable}
\end{table}

%% file: sections/instructions.tex
Section \ref{sec:instructions_main_experiment} presents instructions for the experiment. Analogously, Sections \ref{sec:instructions_svo} and \ref{sec:instructions_bret} present instructions for the social preferences (SVO) and risk preferences (BRET) elicitation tasks respectively.

\subsection{Main Experiment}
\label{sec:instructions_main_experiment}

This section contains instructions for Baseline and Intervention parts of the experiment. Note that the type of network and intervention do not feature in this part of instructions. For the Intervention (Part 2), we show instructions for the fine. Instructions for the nudge are similar, except that instead of explaining how the fine is implemented, participants are asked to watch a 3-minute video. The video can be accessed online at \url{\darija{enter}}. 

\begin{CJK*}{UTF8}{gbsn}

\textbf{第一部分. 界面1：基本信息介绍}

欢迎参加这个互动实验！

该实验包括两个部分。完成全部实验，您将获得¥5美元的固定奖励。此外，您在实验中做的选择还会为您赚取积分，该积分将在本实验结束时转换为美元。有可能还有一额外奖金任务。

在实验的第1部分和第2部分中，您和其他招募到的人会被随机分配组成五人小组，进行交互小游戏。在此过程中，您和其他人的的身份都是保密的，并不知道其他人的身份信息。

实验预期时长为30分钟，您的平均预期总收入为¥13-15（不包括额外奖金任务）。值得注意的是，您的收入可能会低于平均值也可能高于平均值，具体取决于您在实验过程中的选择以及小组中其他人的选择。

在实验的第1部分开始时，您将被随机分配到5人小组，该小组将一直持续到整个实验结束。过程中我们将您与其他4位用户匹配时，您可能需要等待，但是我们会为您等待的时间补偿您相应奖励。

由于此实验是交互式的，因此过程保持持续专注非常重要，否则您可能会拖慢其他人的速度，甚至您自己可能被取消参加实验的资格。

接下来将介绍游戏规则。完成第一部分后，您将收到有关第二部分的更多信息。

所有参与者接收到的指示是相同的。请务必仔细阅读实验导论的说明，我研究方承诺在此实验中并无欺骗参与者行为。

阅读此部分导论后，您将会接受一个简短的理解测验。如果您未通过测验，则将不允许您参加实验，也不会获得固定的奖励

要继续操作说明，请按下面的“下一步”按钮。

\textbf{第一部分. 界面2：实验图示含义介绍}

在实验中，您与小组中的其他成员会一起参与到游戏中，下文中将用“参与者”来称呼您和小组成员。

在游戏开始时，您会看到一个带有5个以大写字母（P，E，C，M，Q）标记的圆圈以及线条相联系的图表。

每个圆圈代表着参与者，该圆圈和线段标注了他和其他参与者的位置关系。在游戏开始时，每个参与者都被随机分配到图中的这些位置上。您的位置是标注成蓝色的圆圈。位置之间的线段表示了参与者之间的位置交互结构。

这些线段表示游戏中哪些参与者是产生交互，即彼此存在接触。

如下图所示，此例子中您处于位置M，可以直接与位置P和E的参与者互动，但不能直接与C和Q互动。

\vspace{0.3cm}
\centerline{\includegraphics[scale=.5]{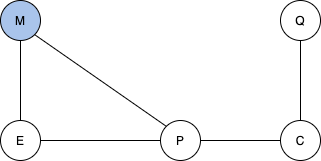}}

下一页操作说明将会解释在此游戏中参与者所做的决定，想要进入下一个界面，请点击“下一页”按钮，想要回到上一个界面，请点击“返回”按钮。

\textbf{第一部分. 界面3：社交距离择界面解释}

在游戏中，您和其他参与者将会面临被新冠病毒(COVID-19)感染的风险。

新冠病毒的主要症状是呼吸急促，高烧和持续的咳嗽。大多数患者会出现轻度症状，并在1-2周内自行恢复，但免疫力弱的个体感染后症状逐步发展为肺炎，并出现器官衰竭。

当你知道游戏中你所处的位置后，你需要做出选择——是否要保持社交距离。下图展示了让您做出选择的按钮的界面外观。
  
\vspace{0.3cm}
\centerline{\includegraphics[scale=1]{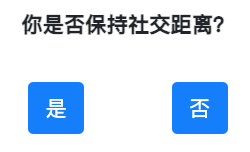}}
  
实验中每个界面的顶部都会有计时器，每一个选项您都有80 秒的时间做出选择。如果您未在允许的时间内做出选择，您的选择会被默认为“否”。

当小组中每个人都做出了选择之后，系统会随机选取游戏中的一名，且只有一名参与者成为新冠病毒的直接感染者。如果参与者没有主动保持社交距离，同时又被系统随机选中成为感染者，那么他/她肯定会受到感染。换句话说，如果参与者不进行社交距离，则有20％的机会直接感染新冠病毒。

社交距离一定程度上为您提供了针对新冠病毒的保护。如果您保持社交距离，此时假如您被系统随机选中感染到新冠病毒，则系统会采取掷硬币的方式决定您是否感染上病毒。如果硬币翻转是正面，则您感染了新冠病毒。如果硬币为反面，则您未感染新冠病毒。换句话说，如果您进行社交距离，则有10％的机会直接感染新冠病毒。

保持社交距离的参与者不会传染新冠病毒病毒给其他参与者，也不会被其他参与者传染新冠病毒。

另一方面，一个感染病毒的参与者如果不选择保持社交距离，则可能会传染其他参与者。特别是，其他没有选择保持社交距离距离的参与者将会面临被感染风险，因为新冠病毒会通过参与者之间的互动而传播。

下一页操作说明将会解释此游戏中新冠病毒是如何在参与者的互动中的传播的，想要进入下一个界面，请点击“下一页”按钮，想要回到上一个界面，请点击“返回”按钮。

\textbf{第一部分. 界面4：新冠病毒传播机制解释}

没保持社交距离的健康者可能会通过与未保持社交距离的感染者接触而感染新冠病毒。这里参与者通过与另一个参与者接触而感染新冠病毒的概率被称为新冠病毒的传染率。

\textbf{在实验的第1部分，新冠病毒的传染率被定为65\%。}

再次考虑界面2中参与者交互示例图，假设：

\vspace{0.3cm}
\centerline{\includegraphics[scale=.5]{images/example_network.png}}

\begin{itemize}
    \item 你（参与者M）没有选择保持社交距离
    \item 在位置E和Q的参与者选择了保持社交距离，而位置P和C上的参与者没有保持社交距离。
\end{itemize}

此时，假设系统随机选定参与者C感染新冠病毒。

首先，C选择不保持社交距离，因此他/他肯定会受到感染。而且，由于C选择不保持社交距离，他也可能将新冠病毒传递给其他参与者。

其次，E和Q选择保持社交距离，因此不会被传染。

接下来，P可能会因新冠病毒传染性而受到感染，因为他/她没有保持社交疏远并且与C存在接触。这种情况的发生概率为65\%（传染率）。

最后，您也可能通过与P的互动而被传染。您与P的互动有65\%的可能性被感染。但是，如果P保持健康，那么您在游戏中也将保持健康。

因此，在以上示例中，如果1）您未保持社交距离，2）E和Q保持社交距离，而P和C未保持社交距离，3）在C被确定为感染者的情况下，则您可能会有42.25\%的概率受到感染。下图显示了概率计算过程。

\vspace{0.3cm}
\centerline{\includegraphics[scale=.75]{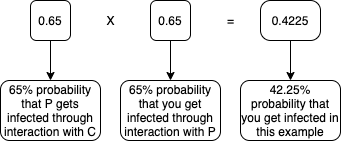}}

注意：您在实验中并不会告知其他参与者的社交距离选择。

下一页操作说明将会解释在第一部分的实验中你是如何赢得积分的。想要进入下一个界面，请点击“下一页”按钮，想要回到上一个界面，请点击“返回”按钮。

\textbf{第一部分. 界面5：实验过程积分赚取说明}

通过在游戏中您“是否保持社交距离”，以及结束时的感染状态，系统会对您进行打分：

\begin{itemize}
    \item \textbf{100分:} 如果您游戏中没有选择保持社交距离，结束时也没有被感染，您将得到100分；
    \item \textbf{65分:} 如果您保持社交疏远，结束时也未受到感染，您将得到65分（100分健康减去35分的社交距离费用）；
    \item \textbf{0分:} 如果您未保持社交疏远，结束时不幸感染，则您将得到0分；
    \item \textbf{-35分:} 如果您保持社交距离，但结果还是被感染，您将被减去35分（0分健康值减去35分的社交距离费用）。
\end{itemize}

请注意，如果您未能在每次游戏开始时提交您的社交距离选择，那么您将减去50积分作为惩罚。

当您选择是否保持社交距离时，有关您积分赚取对应的相应操作结果，新冠病毒的传染率，参与者之间的互动结构以及您的位置信息会始终显示在屏幕上。

如下图所示，您可以看到该步骤界面的样子。交互作用图在右侧，而左侧的文本信息会提醒您感染率以及操作结果对应的积分。

\vspace{0.3cm}
\centerline{\includegraphics[scale=.7]{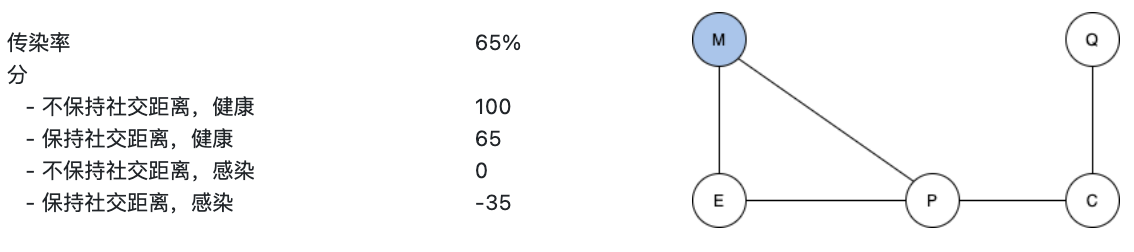}}

游戏结束时，你将会看到的界面会提示您各个参与者的位置结构，您的位置以及您的社交距离选择。同时，你会被告知你的感染状态以及积分赚取的情况。具体如下图所示：

\vspace{0.3cm}
\centerline{\includegraphics[scale=.7]{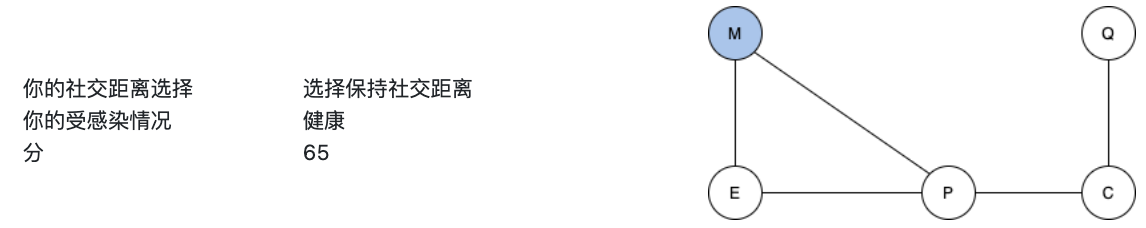}}

此时，你会有20秒去回顾以上信息，计时器会始终置顶于界面顶端。

下一页操作说明将会解释在此游戏所赢得的积分是如何兑换成奖金的，想要进入下一个界面，请点击“下一页”按钮，想要回到上一个界面，请点击“返回”按钮。

\textbf{第一部分. 界面6：积分提现规则}

如本说明所述，实验的第1部分有20个单独的游戏。每个小游戏结果都是独立的，您在一个游戏中所做的选择不会影响其他游戏。

每个游戏的不同之处在于每一次游戏，交互图中参与者的位置是随机分配的。而分配到你所在小组中的参与者，位置间的交互结构，直接感染新冠病毒的几率，新冠病毒的传染率以及由社交距离选择与感染状态对应的点数保持不变。

在每局游戏结束时，您可以查看近5局游戏的选择历史和结果。如下表。

\vspace{0.3cm}
\centerline{\includegraphics[scale=.7]{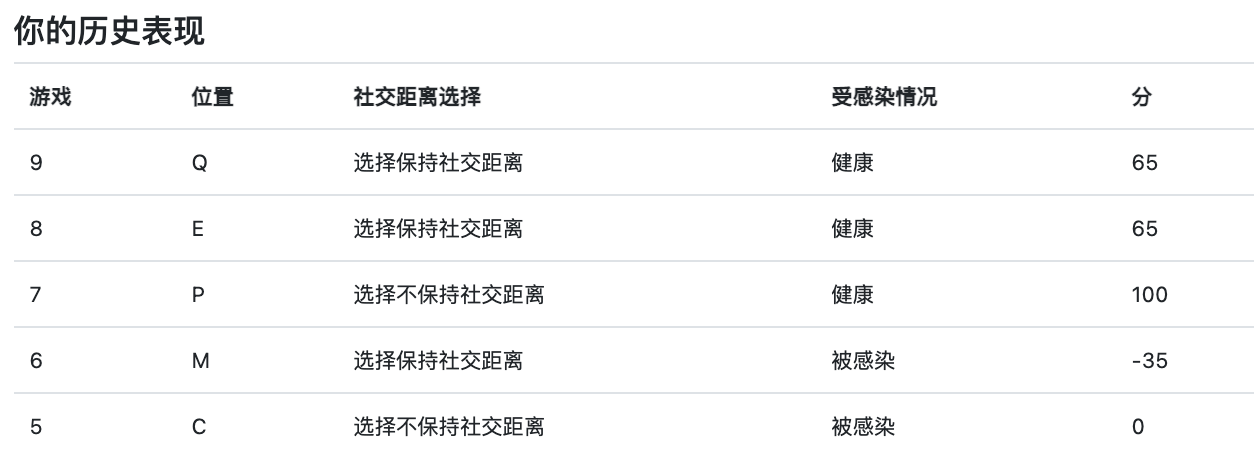}}

请注意：每个游戏的选项您都必须做出选择，如果您连续3个游戏无法做出选择，您将被取消游戏资格。在这种情况下，您将不会得到任何报酬。

在实验结束时，系统将从20个游戏中随机选择4个，以确定您在实验第1部分中的收入。

\textbf{积分将按固定的兑换比例转换成现金报酬（每50分可兑换¥1）。}

假设你在随机选取的4轮次中赚取了260积分，你在第二部分中的总报酬即为¥5.2。

想要开始短测试，请点击“下一页”按钮，想要回到上一个操作解释界面，请点击“返回”按钮。
  
\textbf{第二部分。 说明}

你已经完成了实验的第一部分，并即将进入到实验的第二部分。

以下是实验第二部分的说明，这些说明非常重要，请认真阅读。

实验的第二部分包括20轮游戏，你在实验第一部分的分组不会改变，参与者之间的互动模式，直接感染新冠病毒的概率和新冠病毒的传染率与第一部分一致。

\textbf{第一部分和第二部分唯一的不同在于：在任意轮次游戏中，若你不采取社交隔离，将会被罚款15积分。}

因此，在第二部分实验中，你所赚取的积分是：

\begin{itemize}
    \item \textbf{85分:} 如果您游戏中没有选择保持社交距离(您将被处以罚款15元)，结束时也没有被感染，您将得到85分（100健康分减去15罚款分）；
    \item \textbf{65分:} 如果您保持社交距离，结束时也未受到感染，您将得到65分（100分健康减去35分的社交距离费用）；
    \item \textbf{-15分:} 如果您未保持社交距离(您将被处以罚款15元)，结束时不幸感染，则您将得到-15分 （0分健康分减去15罚款分）；
    \item \textbf{-35分:} 如果您保持社交距离，但结果还是被感染，您将被减去35分（0分健康值减去35分的社交距离费用）。
\end{itemize}

请注意，如果您未能在每次游戏开始时提交您的社交距离选择，那么您将减去50积分作为惩罚。

你在第二部分赚取积分的计算方式与第一部分一致。在实验结束时，系统将会随机选取20轮游戏中的4轮，并以所选轮次结果决定你在第二部分实验的报酬。

与第一部分一致，积分将按固定的兑换比例转换成现金报酬（每50分可兑换¥1）。

假设你在随机选取的4轮次中赚取了300积分，你在第二部分中的总报酬即为¥6。

在你开始第二部分实验之前，你必须先完成一个测试，如果你无法正确回答测试问题，你将无法继续参与到第二部分实验。

请点击“下一步”按钮进入我们的简单测试。

\end{CJK*}
  
\subsection{Social Value Orientation (SVO) Slider Measure}
\label{sec:instructions_svo}

This section contains instructions to the Social value Orientation (SVO) task which participants complete as part of the recruitment survey. 

\begin{CJK*}{UTF8}{gbsn}

您已经正确回答前面的2个问题，并被获准参与到接下来的额外奖金任务。

在这个额外的奖金任务中，您将进行一系列关于您和另一名匿名参与者的报酬分配的行为决策选择。您所有的选择都将会是保密的。

总共有6个关于行为决策的选择，这些行为决策之间是互相独立的。在每个行为决策的选择中，您需要根据您的个人偏好，在您和另一名匿名参与者之间对报酬进行分配（单位为分）。在您完成您的行为决策后，您可以在选项下方点击选择该分配方式。您将会看到，您的选择将会影响到您和另一名匿名参与者的实际获得报酬。

这些行为决策并没有对错之分，仅仅是个人风险偏好差异的体现。

每当有50人完成这个额外奖金任务，我们会随机选取其中2名参与者，并根据下列规则支付额外奖金：我们会在6个行为决策中选择1个，并随机选择这2名参与者中的1名，根据他的真实选择支付额外奖金。

比如，假设我们在50名完成者中选择了X和Y两人，我们随机选择了Y和他的第3个行为决策。假设他的第3个行为决策给他自己分配了¥8.5分，给他人分配了¥6.5分，则X和Y分别获得¥6.5分和¥8.5分。

\end{CJK*}

\subsection{Bomb Risk Elicitation Task (BRET)}
\label{sec:instructions_bret}

This section contains instructions to the Bomb Risk Elicitation Task (BRET) which participants complete after the main experiment.

\begin{CJK*}{UTF8}{gbsn}

非常感谢您参与到这个实验！

您现在位于额外奖金任务环节，您将有可能获得额外奖金。

在下一页，你将会看到100个格子。当你点击\textbf{“开始”}按钮，从左上角开始，每秒钟将会有一个格子被收集（被收集的格子将会打上勾）。

\textbf{每多收集一个格子，你将会多赚到¥0.1。}

在100个格子中，有一个格子会有炸弹，收集到带炸弹的格子会让你失去所有奖金。你并不知道炸弹在哪个格子中，你只知道炸弹以相同的概率出现在任一格子。

你的任务是选择何时停止收集格子，然后打开你收集的所有格子。你可以在任何时候点击\textbf{“停止”}按钮，从而停止收集格子。此后，你可以点击\textbf{“打开”}按钮将所有收集的格子打开。需要注意的是，一旦你点击了\textbf{“停止”}按钮，你将不能重新开始收集格子。每个被收集的格子被打开时，将会显示一个美元符号或者一个炸弹标志。

如果所有收集的格子被打开后都没有炸弹标志，则意味着你并没有收集到带有炸弹的格子。在这种情况下，你可以赚到你收集到的格子数的奖金。

如果炸弹标志出现了，这意味着你收集到了带炸弹的格子。在这种情况下，你在本环节的奖金为0。

\end{CJK*}